\documentclass[traditabstract]{aa} 
\usepackage{graphicx}
\usepackage{stfloats}
\usepackage[varg]{txfonts}
\usepackage{natbib}
\usepackage{lscape}
\usepackage{listings}
\usepackage{courier}
\usepackage[normalem]{ulem}
\usepackage{subfig}
\usepackage{breqn}
\usepackage{caption}
\usepackage{longtable}
\bibpunct{(}{)}{;}{a}{}{,} 

\begin{document}

   \title{Spectroscopic analysis of DA white dwarfs with 3D model atmospheres}

   \author{P.-E. Tremblay\inst{1}
          \and
          H.-G. Ludwig\inst{1}
          \and
          M. Steffen\inst{2}
          \and
          B. Freytag\inst{3}
          }

   \institute{Zentrum f\"ur Astronomie der Universit\"at Heidelberg, Landessternwarte, 
            K\"onigstuhl 12, 69117 Heidelberg, Germany\\
             \email{ptremblay@lsw.uni-heidelberg.de,hludwig@lsw.uni-heidelberg.de}
         \and
            Leibniz-Institut f\"ur Astrophysik Potsdam, An der Sternwarte 16, D-14482 Potsdam, Germany\\
             \email{msteffen@aip.de}
         \and
           Centre de Recherche Astrophysique de Lyon, UMR 5574: CNRS, Universit\'e de Lyon,
           \'Ecole Normale Sup\'erieure de Lyon, 46 all\'ee d'Italie, F-69364 Lyon Cedex 07, France \\
              \email{Bernd.Freytag@ens-lyon.fr}
             }

   \date{Received ..; accepted ..}
 
  \abstract {We present the first grid of mean three-dimensional (3D) spectra
    for pure-hydrogen (DA) white dwarfs based on 3D model atmospheres. We use
    CO5BOLD radiation-hydrodynamics 3D simulations instead of the
    mixing-length theory for the treatment of convection. The simulations
    cover the effective temperature range of $6000 < T_{\rm eff}$ (K) $<
    15,000$ and the surface gravity range of $7 < \log g < 9$ where the large
    majority of DAs with a convective atmosphere are located. We rely on
    horizontally averaged 3D structures (over constant Rosseland optical
    depth) to compute $\langle {\rm 3D}\rangle$ spectra. It is demonstrated
    that our $\langle {\rm 3D}\rangle$ spectra can be smoothly connected to
    their 1D counterparts at higher and lower $T_{\rm eff}$ where the 3D
    effects are small. Analytical functions are provided in order to convert
    spectroscopically determined 1D effective temperatures and surface
    gravities to 3D atmospheric parameters. We apply our improved models to
    well studied spectroscopic data sets from the Sloan Digital Sky Survey and
    the White Dwarf Catalog. We confirm that the so-called high-$\log g$
    problem is not present when employing $\langle {\rm 3D}\rangle$ spectra
    and that the issue was caused by inaccuracies in the 1D mixing-length
    approach. The white dwarfs with a radiative and a convective atmosphere
    have derived mean masses that are the same within $\sim$0.01
    $M_{\odot}$, in much better agreement with our understanding of stellar
    evolution. Furthermore, the 3D atmospheric parameters are in better
    agreement with independent $T_{\rm eff}$ and $\log g$ values from
    photometric and parallax measurements.}{}{}{}{}

   \keywords{convection --- hydrodynamics --- line: profiles --- stars: atmospheres --- white dwarfs }

   \titlerunning{Spectroscopic analysis of DA white dwarfs with 3D model atmospheres}
   \authorrunning{Tremblay et al.}
   \maketitle

\section{Introduction}

White dwarfs with hydrogen lines as their main spectral feature represent
about 75\% of all known degenerate stars \citep{ms99,kleinman13}. The spectral
type of these white dwarfs is called DA and most of them have a pure-hydrogen
atmosphere. The atmospheric parameters of DA stars, the effective temperature
and surface gravity ($T_{\rm eff}$ and $\log g$), are most precisely
determined from spectroscopic analyses through a comparison of observed and
predicted Balmer line profiles
\citep{bergeron92,voss09,gianninas11,kleinman13}. While the hydrogen atom and
corresponding opacities are well known, the predicted atmospheric parameters
are highly sensitive to the shape of the higher series members of the Balmer
lines, which are in turn strongly impacted by complex non-ideal effects
\citep{hm88}. It is only recently that a consistent implementation of the
non-ideal effects directly in the Stark broadening calculations was performed
\citep{TB09}. Other aspects that were recently improved include the H-H$_2$,
H-H, and H-H$^+$ broadening of the lower Lyman lines
\citep{allard04,kowalski06}.

The atmospheric parameters, coupled with the mass-radius relation derived from
structure models of cooling white dwarfs, can be used to constrain masses and
ages. This is a fundamental technique in white dwarf research and
astrophysics, such as in studies of galactic clusters and the halo \citep[see,
  e.g.,][]{hansen99,fontaine01,dobbie06,hansen07,kalirai12}. There is still
much interest to improve our understanding of the model atmospheres and the
mass-radius relation of DA stars. In the latter case, a theoretical sequence
\citep{fontaine01,renado10} with a given internal composition is usually
assumed, although the relation varies by 3-5\% whether a thick or thin
hydrogen layer is chosen. The thickness of this layer is currently poorly
constrained and could vary substantially among the different DAs
\citep{fontaine94,TB08,fontaine08,romero12}. The mass-radius relation is also
difficult to constrain from observations mostly due to the lack of accurate
trigonometric parallax measurements \citep{holberg12}. However, it is hoped
that in the next few years, the white dwarfs observed in the {\it Gaia}
mission \citep{gaia} or other dedicated surveys will constrain the mass-radius
relation.

Pure-hydrogen atmospheres become convective for $T_{\rm eff} \lesssim 14,000$
K at $\log g$ = 8, as the recombination of hydrogen causes a significant
increase in the opacity. The convective zone is initially restricted to a thin
portion of the atmosphere around Rosseland optical depth ($\tau_{\rm R}$)
unity, but rapidly reaches deeper layers during the cooling process, even
though the convective zone itself does not impact the cooling rates at that
stage \citep{fontaine01}. In both atmosphere and structure models, convection
has been traditionally treated with the mixing-length theory
\citep[][hereafter MLT]{MLT}. This one-dimensional (1D) theory relies on as
much as five free parameters \citep{ludwig99} that must be adjusted in order
to describe how energy is transported by convection. The ML2/$\alpha$
parameterisation \citep[][where $\alpha$ is the mixing-length to pressure
  scale height ratio]{tassoul90} is typically utilised in the white dwarf
field. The MLT predicts that when DA white dwarfs are cooling, the bottom of
the convective zone reaches a maximum depth of about $M_{\rm H}/M_{\rm tot} =
10^{-6}$ \citep{grootel12}, which corresponds to a very small fraction of the
radius. This limit is attained when the convective zone reaches the growing
degenerate core \citep{lamb75}. It has been suggested that about 15\% of DAs
have thin hydrogen layers ($M_{\rm H}/M_{\rm tot} < 10^{-6}$), and that they
turn into helium-rich atmospheres when the convective zone mixes the helium
and hydrogen layers \citep{shipman72,mix1,sion84,TB08}.

It has long been suspected that the 1D pure-hydrogen model atmospheres are
inadequate to describe cool DA white dwarfs
\citep{bergeron90,koester09,TB10}. For the past 20 years, authors have
systematically found that the spectroscopic mass distribution of white dwarfs
with a radiative atmosphere ($T_{\rm eff} \gtrsim 13,000$~K and cooling age
$\lesssim$ 0.3 Gyr) has a mean mass in the range of 0.56-0.64 $M{_\odot}$
\citep{bergeron92,gianninas11,TB11,kleinman13}. On the other hand, the
apparent average mass of cool DA stars with a convective atmosphere (0.3 $<$
age [Gyr] $<$ 3) is in the range of $\sim$0.75 $M{_\odot}$. Supposing that the
initial mass function for objects within 200 pc of the Sun has not changed
dramatically in this age range, such difference is incompatible with stellar
evolution models. Considering only the fact that our galaxy is older, white
dwarfs formed recently are coming from progenitors that are on average
slightly older compared to degenerate stars formed 3 Gyr ago. However, current
initial-final mass relations \citep{catalan08,renado10} imply that this would
be a very small effect on the mean mass as a function of cooling age.
Furthermore, the masses determined from parallax, photometric, and
gravitational redshift observations do not show evidences of higher masses for
convective objects \citep{koester09,TB10,falcon10}.

CO$^5$BOLD 3D simulations \citep{freytag12} of DA white dwarf atmospheres
using radiation-hydrodynamics to treat convection have recently been
computed. \citet[][hereafter Paper I]{paper1} presented four simulations of
warm convective DAs close to the ZZ Ceti instability strip. Their results
already indicated that 3D simulations predict surface gravities that are
significantly lower than the 1D models. \citet{tremblay13a}, hereafter Paper
II, improved the 3D simulations by updating the equation-of-state and
opacities. They demonstrated that the mean 3D structures are not particularly
sensitive to the numerical parameters, unlike the 1D models which are
sensitive to the MLT parameterisation, hence suggesting that the 3D results
have a better accuracy. They also increased the number of simulations by
computing a sequence of 12 models at $\log g = 8$ covering the range $6000 <
T_{\rm eff}$ (K) $< 13,000$. It was shown that 3D $\log g$ corrections have
roughly the right amplitude and $T_{\rm eff}$ dependence to solve the
high-$\log g$ problem. Furthermore, they compared in detail the 3D and 1D
structures, and to a lesser degree the spectra, in order to understand the 3D
effects. One conclusion is that two aspects that are missing from current 1D
models, namely the convective overshoot and the departure from hydrostatic
equilibrium, impact significantly the mean 3D structures. We have recently
extended the grid of CO$^5$BOLD 3D simulations to DA white dwarfs with $\log g
=$ 7.0, 7.5, 8.5, and 9.0. These models are part of the CIFIST grid
\citep{cifist,caffau11,tremblay13b} along with giant and dwarf simulations
computed by the Paris GEPI group.

In this work, we compute a grid of spectra from our 3D simulations of DA white
dwarfs in order to study the 3D effects on the spectroscopic determination of
the atmospheric parameters. In Sect.~2, we present our grid of 3D simulations
and mean 3D spectra. The Sect.~3 is dedicated to a study of the 3D effects on
the atmospheric parameters, including the presentation of fitting functions to
convert 1D parameters to 3D. In Sect.~4, we review the properties of the Sloan
Digital Sky Survey and White Dwarf Catalog samples by relying on our 3D
models. We also compare our results to independent photometric and parallax
observations. A discussion on the accuracy of the $\langle {\rm 3D}\rangle$
spectra follows in Sect.~5 and we conclude in Sect.~6.

\section{3D model atmospheres}

\begin{figure}[!h]
\begin{center}
\includegraphics[bb=38 172 572 642,width=3.2in]{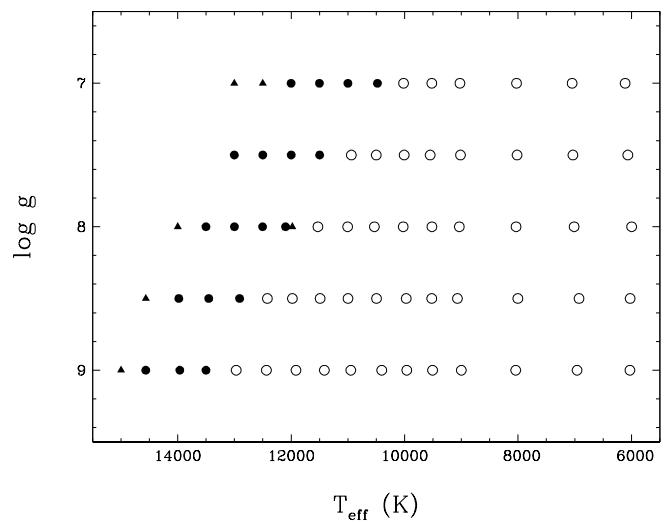}
\caption{Surface gravity and mean $T_{\rm eff}$ for the CO$^{5}$BOLD 3D model
  atmospheres. Simulations were computed with a bottom boundary layer that is open
  (open circles) or closed (filled circles) to convective flows. Models that
  were not selected for the grid of spectra are denoted by a triangle symbol.
\label{fg:f_HR}}
\end{center}
\end{figure}

We rely on 70 simulations of pure-hydrogen white dwarf atmospheres computed
with the CO$^{5}$BOLD code as part of the CIFIST grid. The different
computations are presented in a HR-type diagram in Fig.~\ref{fg:f_HR}. The
numerical setup for the 12 models at $\log g = 8$ and $6000 < T_{\rm eff}$ (K)
$< 13,000$ is described in detail in Paper II and references therein. The
computation of the remaining models has proceeded with the same version of
CO$^{5}$BOLD and the same numerical setup, which we briefly review in this
section. Properties of the individual models, such as $T_{\rm eff}$ (derived
from the temporally and spatially averaged emergent stellar flux), $\log g$,
and computation time, can be found in the online Appendix A.

The implementation of the boundary conditions is described in detail in
\citet[][see Sect.~3.2]{freytag12} and Paper II presents a summary for the case
of our white dwarf models. The hottest simulations, represented by filled
symbols in Fig.~\ref{fg:f_HR}, were computed with a bottom layer that is
closed to convective flows (zero vertical velocities). In those models, the
convection zone is thinner than the typical vertical dimension of the
atmosphere. In cooler models, an open lower boundary is necessary to transport the
convective flux in and out of the domain. We specify the entropy of the
ascending material to obtain approximately the desired $T_{\rm eff}$ value. In
all cases, the lateral boundaries are periodic, and the top boundary is open
to material flows and radiation.

We adopt a grid of $150\times150\times150$ points for all simulations. The top
boundary reaches a space- and time-averaged value of no more than $\log
\tau_{\rm R} \sim -5$ and the bottom layer is generally around $\log \tau_{\rm
  R} \sim 3$. The number of pressure scale heights covered by the simulations
between the photosphere ($\tau_{\rm R} = 1$) and the open bottom boundary is
generally higher than 3 (see online Appendix~A), which ensures that convective
eddies reaching the photosphere are unlikely to be impacted by boundary
conditions. For models with a closed bottom boundary, we have ensured that the
position of the boundary is about 1 dex in $\log \tau_{\rm R}$ below the
convectively unstable region to have a proper account of the overshoot
region. The horizontal geometrical dimensions were chosen so that of the order
of $3\times3$ granules are included in the simulations. In Appendix A, we
provide the box dimensions along with the characteristic granulation size (see
\citealt{tremblay13b} for the derivation).

We use EOS and opacity tables that rely on the same microphysics as the 1D
models of \citet{TB11}. In brief, we employ the \citet{hm88} EOS, the Stark
broadening profiles of \citet{TB09} and the quasi-molecular line opacity of
\citet{allard04}. We use band-averaged opacities to describe the
band-integrated radiative transfer, based on the procedure laid out in
\citet{nordlund82,ludwig94} and \citet{voegler04}.

The wavelength-dependent opacities are sorted based on the Rosseland optical
depth at which \mbox{$\tau_\lambda = 1$}. We employ thresholds in $\log
\tau_{\rm R}$ given by [$\infty$, 0.0, $-0.5$, $-1.0$, $-2.0$, $-3.0$, $-4.0$, $-\infty$]
for the 8 bin setup and [$\infty$, 0.25, 0.0, $-0.25$, $-0.5$, $-1.0$, $-1.5$, $-2.0$,
  $-3.0$, $-4.0$, $-\infty$] in the case of the new 11 bin configuration discussed
below. In both contexts, we added one bin for the Lyman quasi-molecular
satellites. The different opacity tables were sorted by means of reference 1D
model atmospheres of the same $\log g$ as the 3D simulations and $\Delta T_{\rm
  eff} <$ 1000~K.

We demonstrated in Paper II (see Sect.~2) that a total of 8 opacity bins was
sufficient to reproduce with a good accuracy the monochromatic radiative
energy exchanges in white dwarfs at 8000~K and 12,000~K. In this work, we have
found that for $T_{\rm eff} \ge 12,000~$K, where UV flux and opacities become
dominant in the photosphere, a 11 opacity bin approach produced slightly
better results. In particular, the agreement is better between 1D hydrostatic
structures relying on the opacity binning procedure \citep{caffau07}, and the
\citet{TB11} white dwarf models with 1813 carefully chosen frequencies for the
radiative transfer. The differences are rather small ($<$1\%) in terms of the
predicted equivalent width of the Balmer lines. However, it is important that
the warm 3D simulations, with small 3D effects, are precise enough so that we
can connect them with the grid of 1D models. Furthermore, it is shown in this
work that 3D $T_{\rm eff}$ corrections for $T_{\rm eff} > 12,000$~K are rather
sensitive to the predicted structures. We have therefore recomputed all models
with a closed bottom boundary by utilising 11 opacity bins.

To make sure that our simulations have relaxed in the upper layers, we
performed non-grey 2D simulations for cooler models with long radiative
relaxation timescales (see Paper II, Sect.~3.3). We have found that the upper
layers never actually reach a radiative equilibrium like the 1D
models. Instead the convective overshoot causes the entropy gradient in the
upper layers to relax to a near-adiabatic structure. Once our 2D models
reached a nearly adiabatic structure, they were used as initial conditions for
3D simulations.

All non-grey 3D simulations were run long enough to cover typically $\sim$100
turnover timescales in the photosphere. Computation times are given in
Appendix A, and we refer to \citet{tremblay13b} for a discussion about the
characteristic timescales of the granulation. We have verified that all models
are relaxed in the second half of simulation runs and that they show no
systematic (non-oscillatory) change of their properties on timescales longer
than the turnover timescale.

A detailed discussion of the mean 3D structure properties at $\log g = 8$ and
their comparison to standard 1D model atmospheres is found in Paper
II. Following this analysis, \citet{tremblay13b} reviewed the properties of
the surface granulation for 60 models from our grid at different $\log g$. The
authors have shown that the amplitude of the temperature fluctuations, one
measure of the strength of the 3D effects, is well correlated with the density
at $\tau_{\rm R} = 1$. Since the atmospheric density can be kept roughly
constant by increasing both $\log g$ and $T_{\rm eff}$, it was demonstrated
that sequences of 3D simulations at different gravities are fairly similar,
but with a shift in $T_{\rm eff}$ in terms of their granulation
properties. The comparison of mean 3D and 1D structures at different gravities
indeed reveals properties that are analogous to those found in Paper II at
$\log g = 8$. Hence, we do not discuss further the properties of the 3D
structures. In the following, we restrict our study to the 3D effects on the
predicted spectroscopic atmospheric parameters.

\subsection{Model spectra}

Our goal is to apply the 3D model atmospheres to spectroscopic analyses, hence
the first step is to compute 3D spectra. By employing the Linfor3D
three-dimensional spectral synthesis code \citep{linfor}, it was determined in
Paper I that normalised H$\beta$ spectra computed from properly averaged
$\langle {\rm 3D}\rangle$ structures, hereafter $\langle {\rm 3D}\rangle$
spectra, were nearly identical to the results of a full 3D spectral
synthesis. We find that this behaviour is observed for all $T_{\rm eff}$ and
$\log g$ values in our grid. The spatial and temporal averages are performed
over surfaces of constant Rosseland optical depth and for 12 random snapshots.

\begin{figure}[!h]
\begin{center}
\includegraphics[bb=38 172 572 642,width=3.2in]{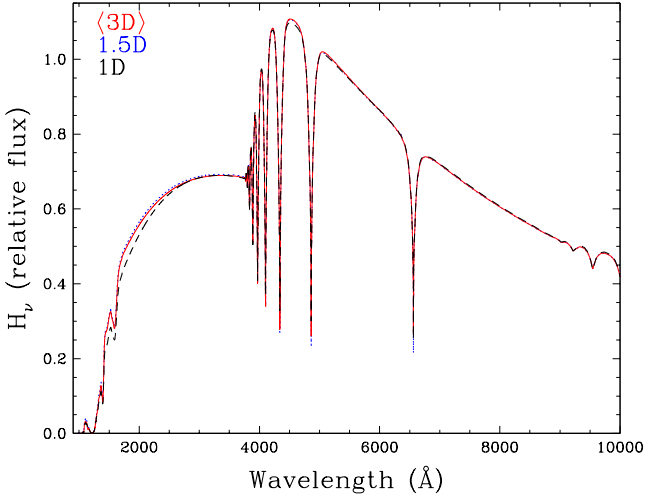}
\caption{Predicted flux at $T_{\rm eff} = 12,022$~K and $\log g = 8$ based on
  a $\langle {\rm 3D}\rangle$ structure (solid, red), the 1.5D approximation
  (see text; blue, dotted) and a 1D ML2/$\alpha$ = 0.8 structure (black,
  dashed). All spectra were normalised at 4760~\AA.
\label{fg:f_flux}}
\end{center}
\end{figure}

To further constrain the precision of the $\langle {\rm 3D}\rangle$ approach,
we have computed a spectrum in Fig. ~\ref{fg:f_flux} with the so-called 1.5D
approximation \citep{steffen95}. This method consists of computing the
emergent flux at the top of each grid point of the 3D simulation, assuming
that the physical conditions do not vary in the horizontal direction,
i.e. each columns is a 1D (plane-parallel) atmosphere. The 1.5D spectrum is
then the average over all individual spectra. The full 3D spectral synthesis
has the effect of coupling nearby grid points, hence it is expected to lie
somewhere between the extreme cases of the 1.5D and $\langle {\rm 3D}\rangle$
approximations. Our experiment demonstrates that the predicted 1.5D flux is
nearly identical to the $\langle {\rm 3D}\rangle$ flux in all wavelength
ranges (see Fig.~\ref{fg:f_flux}), including the hydrogen lines. It confirms
the result of Paper I that a single $\langle {\rm 3D}\rangle$ structure is
adequate for high precision spectral synthesis\footnote{These results apply to
  the normalised flux and 3D effects are slightly larger in terms of the
  absolute flux.}.

It is not straightforward to explain the physical reasons behind the
correspondence between 3D and $\langle {\rm 3D}\rangle$ spectra. \citet{ku13}
studied 3D vs. $\langle {\rm 3D}\rangle$ effects for metal lines, but these
results can not be applied to non-trace elements and saturated lines. In
principle, the monochromatic flux would be best represented if $T^4$ was
averaged over surfaces of constant $\tau_{ \lambda}$ instead of one unique
$\tau_{\rm R}$ scale. However, our results demonstrate that the mean
monochromatic flux is well described by a single mean structure, which we
traced back to the fact that the average over constant $\tau_{\rm R}$ provides
a very good approximation of the mean monochromatic source function in the
intensity forming layer

\begin{equation}
\langle S_{\lambda}(\tau) \rangle_{\tau_{\lambda}=1} \sim \langle S_{\lambda}(\tau)
\rangle_{\tau_{\rm R}} (\langle \tau_{\lambda} \rangle = 1 )~,
\end{equation}

{\noindent}with

\begin{equation}
{\rm d}\langle \tau_{\lambda} \rangle = - \langle \rho \rangle_{\tau_{\rm R}}
\kappa_{\lambda}(\langle T^4 \rangle_{\tau_{\rm R}}^{1/4},\langle P
\rangle_{\tau_{\rm R}}) {\rm d}\langle z \rangle_{\tau_{\rm R}}~,
\end{equation}

{\noindent}where $\rho$ is the density, $\kappa$ the opacity per gram and $P$
the pressure. We suggest that this behaviour is largely caused by a much more
rapid variation of $\tau_{\rm R}$ as a function of geometrical depth in
comparison to temperature. Hence, even if the $\tau_{\rm R}$ scale might not
provide the best averaging surface, the resulting error on the mean
monochromatic flux is very small for white dwarfs. We hope to develop this
result more generally for all stellar types in a future work.

In terms of the temporal average, \citet[][see Eq. (56) and (59)]{l06}
demonstrated that relative temporal variations of the box-averaged emergent
flux scale approximately as

\begin{equation}
\frac{\sigma_f}{\langle f \rangle} \sim 0.5 N_{\rm
  snapshot}^{-1/2} \frac{\delta I_{\rm rms}}{\langle I \rangle} \frac{l_{\rm gran}}{l_{\rm box}}~,
\end{equation}

{\noindent}where $\delta I_{\rm rms}/{\langle I \rangle}$ is the spatial
relative intensity contrast (see Eq.~(5) of \citealt{tremblay13a}) and $l_{\rm
  gran}/l_{\rm box}$ the linear dimension of one granule in box size
units. The factor of $\sim$0.5 comes in part from the center-to-limb darkening
and the fact that intensity is spatially correlated in granules. Given our
simulations maximum intensity contrast of $\sim$20$\%$ (see Appendix~A), it
implies that the temporal flux variation is always less than 1\% when we
average 12 snapshots. Since we rely on normalised line profiles and not on the
absolute flux, the impact on the predicted atmospheric parameters is
significantly less than 1\%.

The results of this section allow us, as in Paper II, to rely on the 1D
spectral synthesis (and model atmosphere) code of \citet{TB11} with the same
microphysics as in our 3D simulations. We use $\langle T^4 \rangle$ and
$\langle P \rangle$ structures from 3D simulations as input to compute
$\langle {\rm 3D}\rangle$ spectra, although populations and monochromatic
opacities are recalculated in the spectral synthesis code. We also employ
through this work 1D pure-hydrogen model spectra computed with the same
code. These models are based, unless otherwise noted, on structures adopting
the ML2/$\alpha$ = 0.8 parameterisation of the MLT. This calibration is
established from a comparison of near-UV and optically determined $T_{\rm
  eff}$ \citep{TB10}, which we review in Sect.~3.3.

In Fig.~\ref{fg:f_flux}, we compare normalised 1D and $\langle {\rm
  3D}\rangle$ spectra at $T_{\rm eff} \sim 12,000$~K and $\log g = 8$. The
Fig.~16 of Paper II further highlights the differences in the wings of the
Balmer lines. Clearly, the differences are fairly subtle and are mostly
related to the shape of the line profiles. We have verified that there is no
significant change in the predicted colours in the optical region. The flux in
the wing of the Ly-$\alpha$ line, however, can be as much as 15\% lower in the
1D case, although this depends on the employed MLT parameterisation. On the
other hand, the absolute fluxes (not shown on the figure) are slightly offset
due to 3D vs. $\langle {\rm 3D}\rangle$ effects, temporal averages, and the
different radiative transfer routines in the 1D spectral synthesis and the 3D
simulations.

\section{Atmospheric parameters}

\begin{figure}[!]
\captionsetup[subfigure]{labelformat=empty}
\begin{center}
\subfloat[]{
\includegraphics[bb=98 189 498 622,width=2.0in]{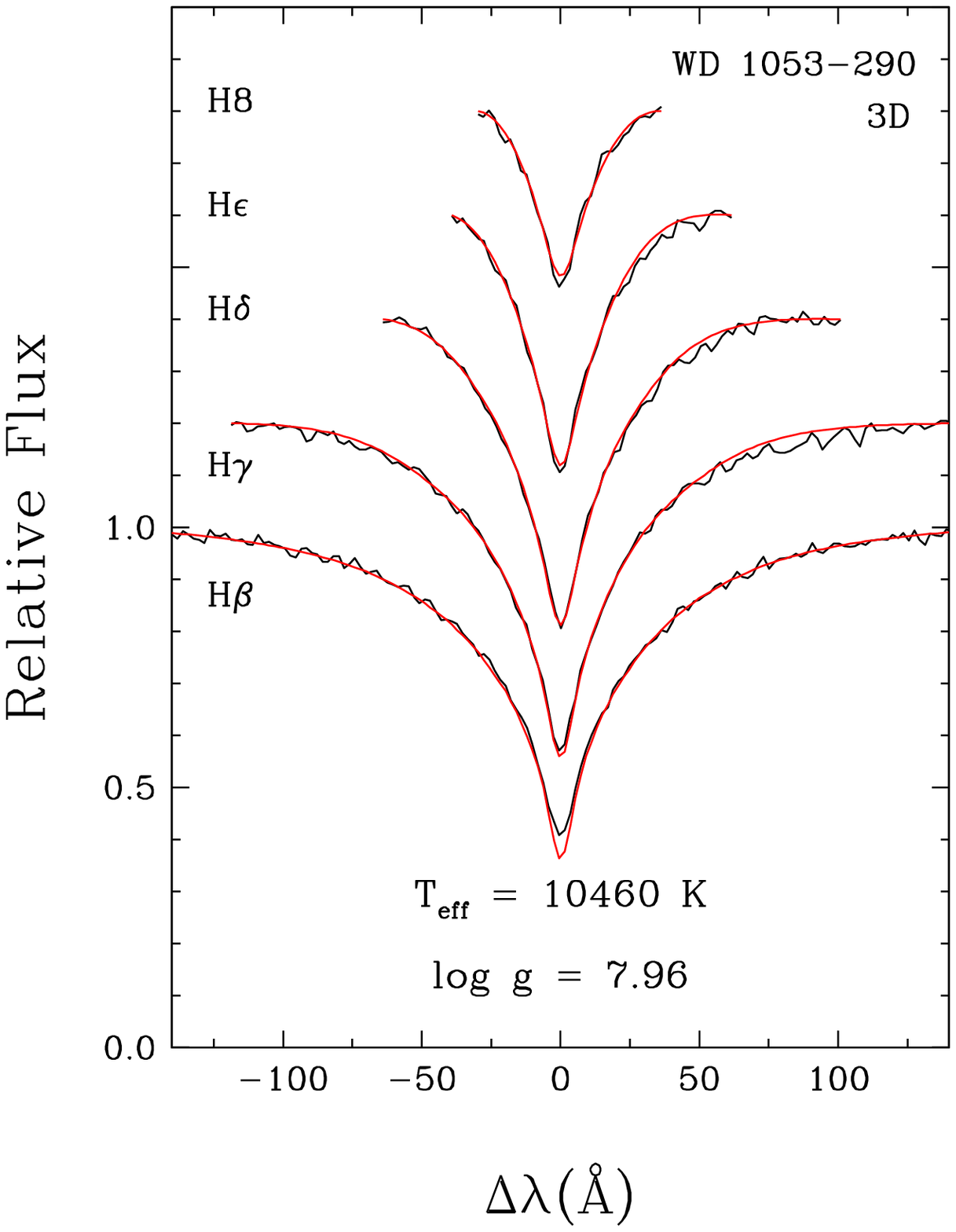}}
\subfloat[]{
\includegraphics[bb=198 189 598 622,width=2.0in]{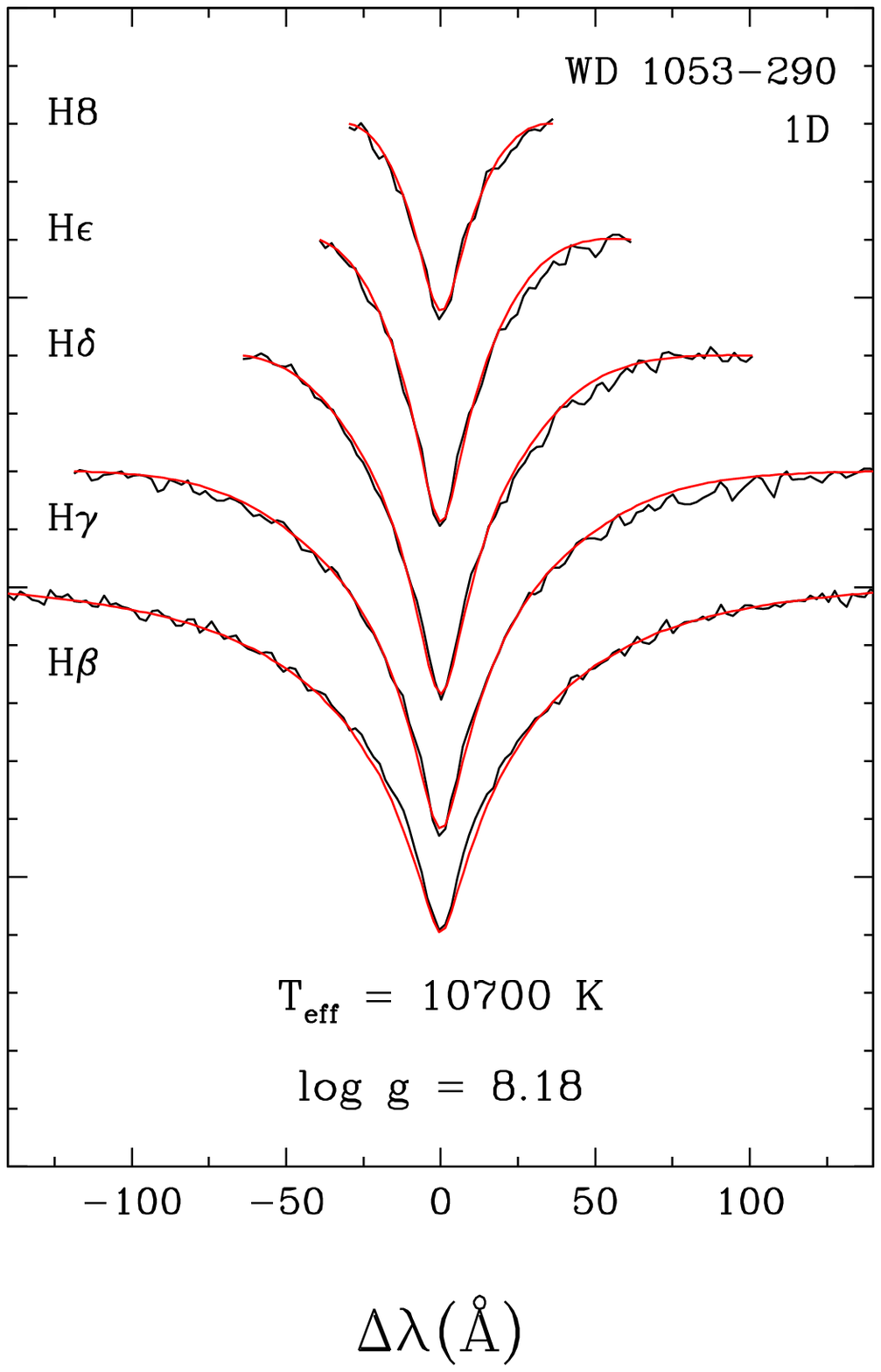}}
\caption{Simultaneous fit of the Balmer lines, from H$\beta$ to H8, for WD
  1053$-$290 from the sample of \citet{gianninas11} by relying on $\langle
  {\rm 3D}\rangle$ (left panel) and 1D ML2/$\alpha$ = 0.8 (right panel) model
  spectra. Line profiles are offset vertically from each other for clarity and
  the best-fit atmospheric parameters are identified at the bottom of the
  panels. The instrumental resolution is of 6~\AA. Line cores were partially
  removed from the $\langle {\rm 3D}\rangle$ fits (see Eq.~(4)).
\label{fg:f_fits}}
\end{center}
\end{figure}

The method used to derive atmospheric parameters from observed line profiles
relies on the so-called spectroscopic technique developed by
\citet{bergeron92}. The first step is to normalise the flux from each
individual Balmer line, typically from H$\alpha$ or H$\beta$ to H8, in both
observed and model spectra, to a continuum set to unity at a fixed distance
from the line center. The comparison with model spectra, which are convolved
with the appropriate Gaussian instrumental profile, is then for these line
profiles only. We rely on the same technique as the one described in
\citet{gianninas11} to define the continuum of the observed spectra. In
Fig.~\ref{fg:f_fits}, we present an example of our fitting procedure for one
cool star in the \citet{gianninas11} sample (see Sect.~4.1) with both the
$\langle {\rm 3D}\rangle$ and 1D grid. The following sections provide a more
detailed discussion of the differences between $\langle {\rm 3D}\rangle$ and
1D fits.

As a first step before using the $\langle {\rm 3D}\rangle$ spectra in
spectroscopic analyses, we have to combine the $\langle {\rm 3D}\rangle$ grid
with hotter and cooler 1D models since our 3D calculations do not cover the
full range of $T_{\rm eff}$ for DA stars. We have created a combined grid of
$\langle {\rm 3D}\rangle$ and 1D spectra in the range $6000 < T_{\rm eff}$ (K)
$< 140,000 $ and $\log g$ = 7.0, 7.5, 8.0, 8.5, and 9.0. We rely on $\langle
{\rm 3D}\rangle$ spectra for $T_{\rm eff}$ up to 12,000, 12,500, 13,500,
14,000, and 14,500~K, for $\log g$ = 7.0, 7.5, 8.0, 8.5, and 9.0,
respectively. The transition corresponds roughly to the position of the
maximum strength of the H$\beta$ line. The next hotter model in the grid is a 500~K
warmer 1D model. The justification for this transition is explained in
Sect.~3.3. We note that the 1D models are identical to those used in
\citet{TB11}, e.g. spectra derived from NLTE TLUSTY structures
\citep{hubeny95} are used for $T_{\rm eff} >$ 40,000~K.

\begin{figure}[!h]
\begin{center}
\includegraphics[bb=38 172 572 642,width=3.2in]{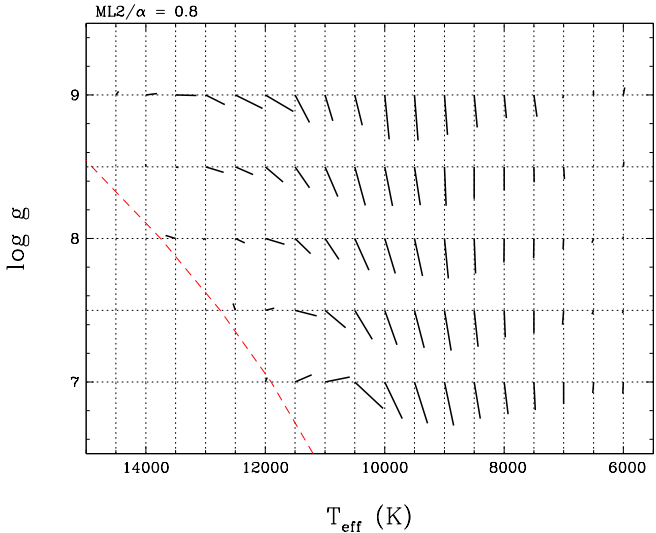}
\caption{3D atmospheric parameter corrections found by fitting our grid of
  $\langle$3D$\rangle$ spectra with the reference grid of 1D spectra relying
  on the ML2/$\alpha$ = 0.8 parameterisation of the MLT. The 1D = 3D reference
  parameters are on the intersection of the dotted lines, and 3D corrections
  are read by following the solid lines. We utilised a resolution of 3~\AA~and
  the cores of the deeper lines were removed from the fits (see Eq.~(4)). The
  red dashed line represents the position of the maximum strength of the
  H$\beta$ line in the 1D models. Tabulated values are available in the online
  Appendix B.
\label{fg:f_shifts}}
\end{center}
\end{figure}

Fig.~\ref{fg:f_shifts} presents the 3D atmospheric parameter corrections found
by fitting the $\langle {\rm 3D}\rangle$ spectra with our standard grid of 1D
spectra. The 3D corrections were derived simultaneously for the five Balmer
lines from H$\beta$ to H8 in the same way we fit observations. The line cores
were partially removed from the fits, as well as the entire H$\alpha$ line,
which is further discussed in Sect.~3.1. In the following sections, we study
the 3D effects in more detail.

\subsection{Line cores}

The $\langle {\rm 3D}\rangle$ structures deviate significantly from their 1D
counterparts in the upper layers ($\tau_{\rm R} < 10^{-2}$) due to the cooling
effect of convective overshoot. As demonstrated in Paper II, this results in
deeper Balmer line cores for the $\langle {\rm 3D}\rangle$ spectra in
comparison to the 1D case. The cooling effect is less significant for our
hottest 3D simulations likely because the radiation field is stronger,
although the impact on Balmer line cores is still substantial because the
lines have their maximum strength near 13,500~K at $\log g = 8$.

The $\langle {\rm 3D}\rangle$ line cores cause potential problems for
spectroscopic analyses. First of all, it is not straightforward to combine the
$\langle {\rm 3D}\rangle$ and 1D spectral grid since even the hottest 3D
simulations still deviate from the 1D structures. Secondly, the predicted line
cores of H$\alpha$ and H$\beta$ are systematically too deep compared to
observations. For the typical case presented in Fig.~\ref{fg:f_fits}, it is
observed that while the overall quality of the fit is rather similar for the
$\langle {\rm 3D}\rangle$ and 1D models, the H$\beta$ line core is too deep in
the 3D case. The problem is more in evidence when the H$\alpha$ line is
included.

We have no indication that our 3D simulations are inaccurate, and a similar
dynamic cooling effect is predicted from 3D simulations of metal-poor dwarfs
and giants computed with different codes \citep{cooling1,cooling2,cooling3}.
In Paper II we have mentioned that missing NLTE effects are unlikely to cause
a discrepancy in the line cores. The problem is also unlikely to be a direct
effect of the limited number of opacity bins or the averaging procedure for
$\langle {\rm 3D}\rangle$ structures. However, it is not excluded that the
significant and simultaneous increase of the resolution, computation time, and
number of opacity bins would lead to conditions of radiative equilibrium in
the uppermost layers where the line cores are formed. Alternatively, the
agreement of 1D spectra with observations could be a coincidence, implying
that some physics is still missing from the models, e.g. magnetic fields
amplified by convection or a proper line broadening theory for the line cores.

We have reviewed different ways of treating the line cores in the $\chi ^2$
fitting procedure and found that the impact on 3D $\log g$ corrections is
small. Best-fit surface gravities vary by less than $\sim$$\vert 0.04 \vert$
dex when a significant part of the line centers is removed, compared to
overall 3D $\log g$ corrections reaching $\sim$$\vert 0.20 \vert$ dex. We
found that the strongest effect of the cores is to shift $T_{\rm eff}$ values
at the hot end of the 3D simulation sequence, near the position of the
maximum strength of the Balmer lines. This is problematic in terms of
combining the grid of $\langle {\rm 3D}\rangle$ spectra with hotter 1D models
and it implies that the accuracy of 3D temperature determinations for $T_{\rm
  eff} \gtrsim 12,500$ K is only of the order of $\sim$250~K. For cooler
models, 3D $T_{\rm eff}$ corrections are smaller, and become less dependent on
the line cores.

In light of these experiments, we rely on the following parameterisation to
remove the line centers from our $\langle {\rm 3D}\rangle$ fits. The line
centers are removed when the following two conditions are verified.

\begin{equation}
\vert \Delta \lambda \vert~< 1.0~\AA ~~{\rm and}~~ \frac{F_{\nu}}{F_{
  \nu,~\rm continuum}} = F_{\nu,~\rm normalised} < 0.6
\end{equation}

{\noindent}The latter requirement ensures that weak Balmer lines, such as
higher lines or even the lower lines in very cool objects, are still taken
into account. Furthermore, we have decided to neglect H$\alpha$ altogether.
The line core problem is the most important for this line, and since it is
situated in a different wavelength region with a different continuum flux, we
find that it is better to remove the line altogether than use a more complex
parameterisation to remove line cores. We employ Eq.~(4) through this work
except when we fit observed Balmer lines with 1D spectra in order to match
published results.

\begin{figure*}[!]
\captionsetup[subfigure]{labelformat=empty}
\begin{center}
\subfloat[]{
\includegraphics[bb=38 189 572 642,width=3.2in]{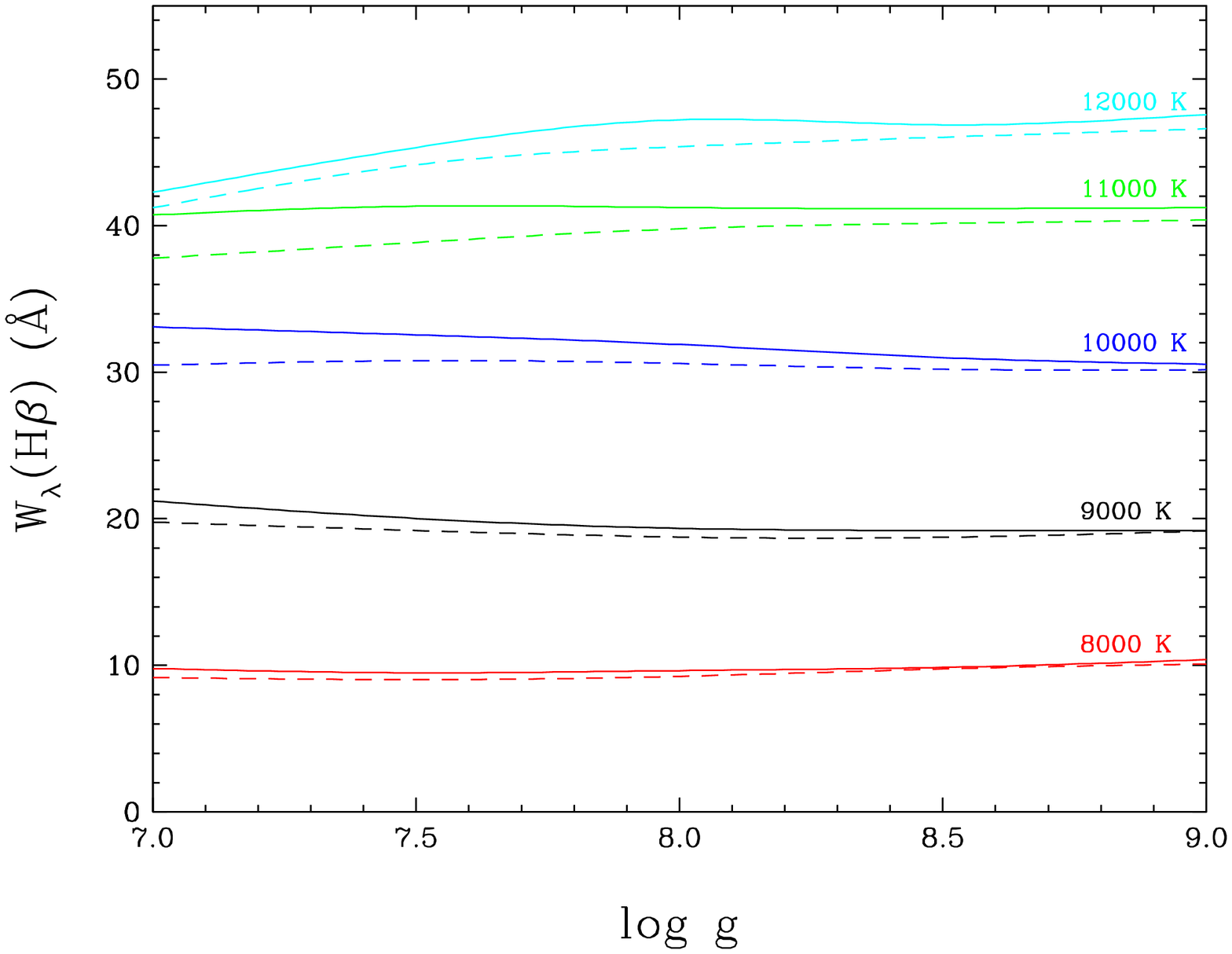}}
\hspace{8mm}
\subfloat[]{
\includegraphics[bb=38 189 572 642,width=3.2in]{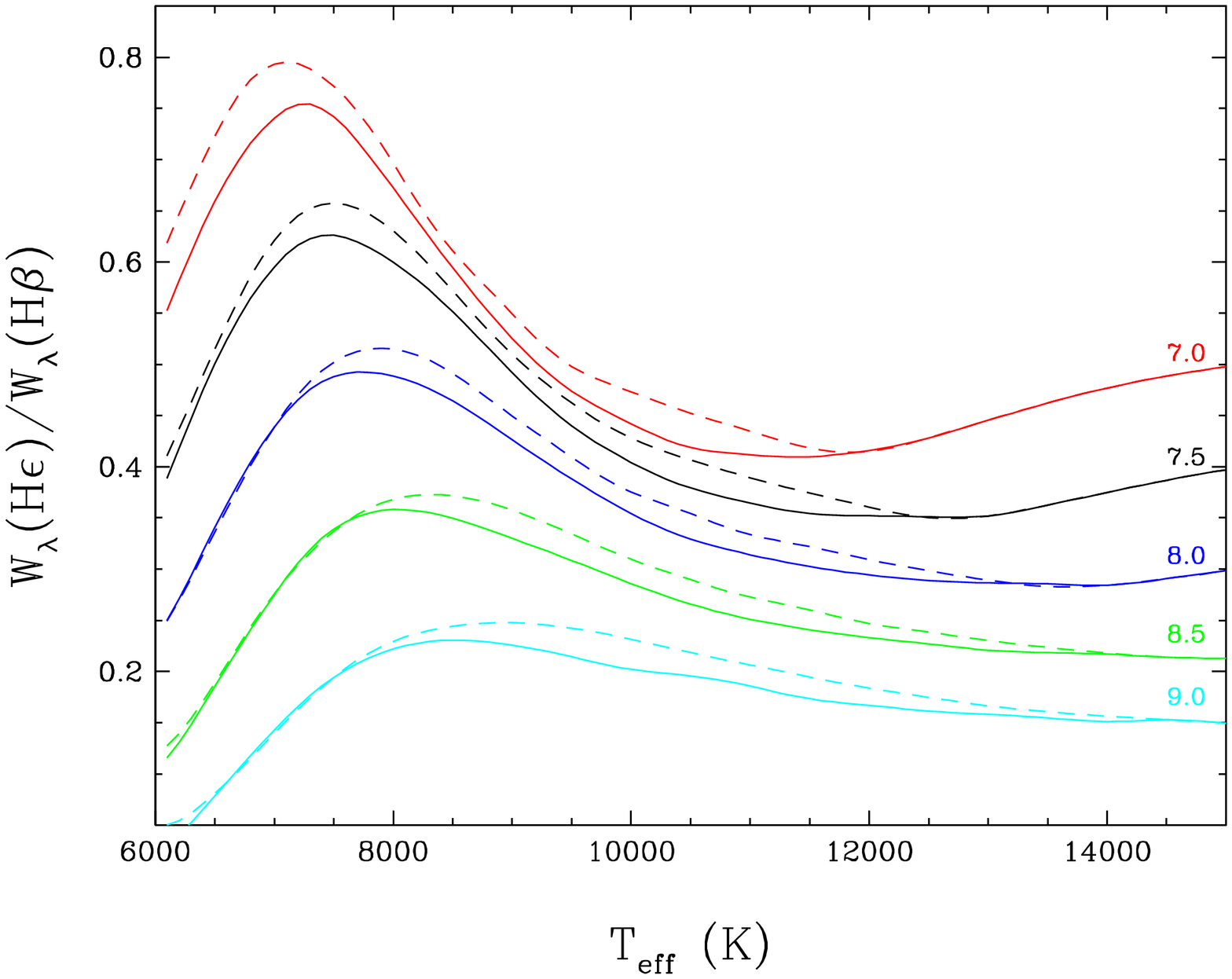}}
\caption{{\it Left:} Equivalent width of the H$\beta$ line for
  $\langle$3D$\rangle$ (solid line) and 1D (dashed) spectra as a function of
  $\log g$ for $T_{\rm eff}$ = 8000, 9000, 10,000, 11,000, and 12,000~K
  (values identified on the panel). {\it Right:} Ratio of the H$\epsilon$ and
  H$\beta$ equivalent widths as a function of $T_{\rm eff}$ for $\log g$ =
  7.0, 7.5, 8.0, 8.5 and 9.0 (values identified on the panel). The line cores
  are not included in the equivalent width (see Eq.~(4)).
\label{fg:f_equiv}}
\end{center}
\end{figure*}

\subsection{3D effects}

It was shown in Paper II that the differences are fairly subtle between the
predicted $\langle {\rm 3D}\rangle$ and 1D spectra, and from that alone it is
difficult to explain the 3D effects on the atmospheric parameters. From the
comparison of the $\langle {\rm 3D}\rangle$ and 1D fits in
Fig.~\ref{fg:f_fits}, it is found that the fit quality is very similar,
even though the best fit surface gravities are fairly different. Nevertheless,
Fig.~\ref{fg:f_shifts} demonstrates that 3D corrections show a well defined
pattern in the HR diagram which we want to explain in this section. 3D
corrections are mostly a function of $T_{\rm eff}$ with little influence from
$\log g$ except for the hottest convective models.

Fig.~\ref{fg:f_equiv} (left) presents the equivalent width of H$\beta$ as a
function of $\log g$ for different $T_{\rm eff}$ values. It demonstrates that
the strength of H$\beta$ is very sensitive to $T_{\rm eff}$, but is rather
independent of $\log g$. The lower lines are therefore largely $T_{\rm eff}$
indicators in the convective regime. As a consequence, the convective
overshoot cooling effect influencing the shape of the line cores for the lower
Balmer lines mostly has an impact on 3D $T_{\rm eff}$ corrections.

On the right panel of Fig.~\ref{fg:f_equiv}, we show the ratio of the
H$\epsilon$ and H$\beta$ equivalent widths as a function of $T_{\rm eff}$ for
different surface gravities. It demonstrates that this ratio is rather
sensitive to $\log g$ and to a lesser degree $T_{\rm eff}$. Furthermore, there
are significant differences between the 3D and 1D cases, which largely
explains the strong 3D $\log g$ corrections in Fig.~\ref{fg:f_shifts}. This is
in agreement with the observation from Paper II that higher Balmer lines are
more significantly impacted by 3D effects than lower lines. The reason for
this behaviour is that the strength of the higher lines is more sensitive to
the non-ideal effects \citep{hm88,TB09}. These effects are in turn responsive
to the density, and $\langle {\rm 3D}\rangle$ structures have systematically
lower temperatures and higher densities in the formation region of the higher
lines in the range $0.1 < \tau_{\rm R} < 1$ (see Fig.~7 of Paper II). Since
increasing the surface gravity of a model also enhances the photospheric
density, 3D effects are largely negative $\log g$ corrections.

The underlying reasons explaining the differences between $\langle {\rm
  3D}\rangle$ and 1D structures in the formation region of the higher lines
appear to be complex. The smaller temperature gradient in the 1D structures
are caused by the fact that the 1D ML2/$\alpha$ = 0.8 parameterisation is more
efficient than the 3D convection in that region. An explanation for this
behaviour may not be possible until we have a general method to improve the 1D
models to make them in close agreement with the $\langle {\rm 3D}\rangle$
structures. It is clear, however, that the 3D convective flux has a smoother
profile as a function of depth in this transition region below the convective
overshoot layer, while the 1D models have a rather sharp transition from
convective to radiative conditions (see Fig.~9 of Paper II).

At the cool end of the sequence, both 1D and 3D structures are nearly fully
adiabatic with a flat entropy profile from the bottom boundary and up to
$\tau_{\rm R} \sim 10^{-2}$. While the 1D structures reach radiative
equilibrium in the upper layers, 3D structures remain adiabatic (see Fig.~8 of
Paper II).  Since the 1D and 3D models are based on the same EOS, the
adiabatic temperature structure is nearly identical in both cases. The
hydrogen lines are fairly weak in this regime, hence the upper layers
($\tau_{\rm R} < 10^{-2}$) have little effect on the predicted flux, and the
3D atmospheric parameter corrections are very small.

\subsection{Calibration of the mixing-length theory}

The 3D atmospheric parameter corrections in Fig.~\ref{fg:f_shifts} are
attached to the particular 1D reference grid, hence to the calibration of the
choice of the MLT parameterisation. While \citet{TB10} suggest to rely on the
ML2/$\alpha$ = 0.8 parameterisation, our results demonstrate that these 1D
models are still significantly different to the 3D predictions, especially in
terms of $\log g$ corrections.  Recent analyses have employed 1D models based
on the ML2/$\alpha$ = 0.6 parameterisation \citep[see,
  e.g.,][]{voss09,kleinman13}, hence it is pertinent to compare the $\langle
{\rm 3D}\rangle$ spectra with 1D models involving different MLT calibrations.

\begin{figure}[!]
\captionsetup[subfigure]{labelformat=empty}
\begin{center}
\subfloat[]{
\includegraphics[bb=38 282 572 662,width=3.2in]{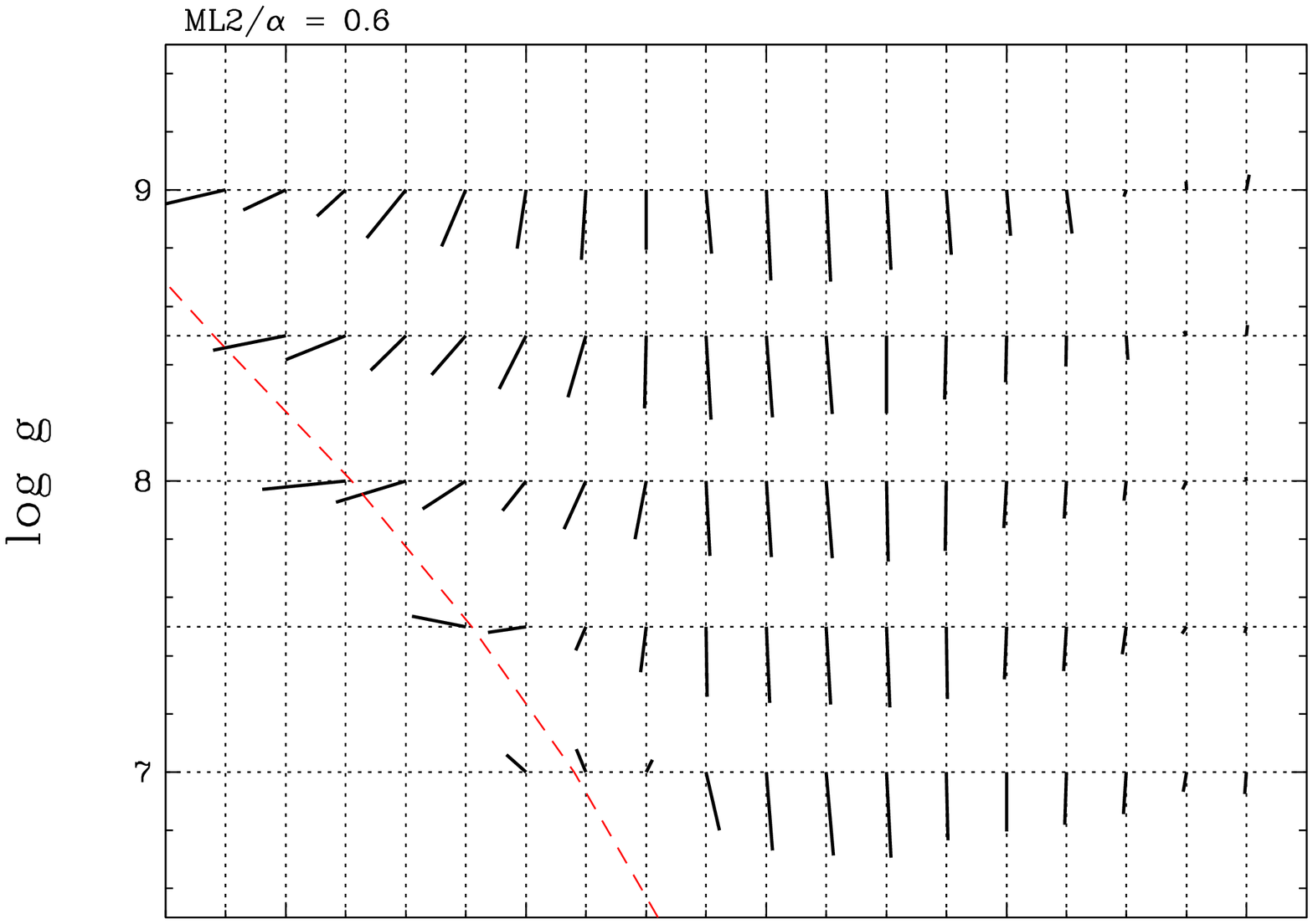}}\\
\subfloat[]{
\includegraphics[bb=38 195 572 642,width=3.2in]{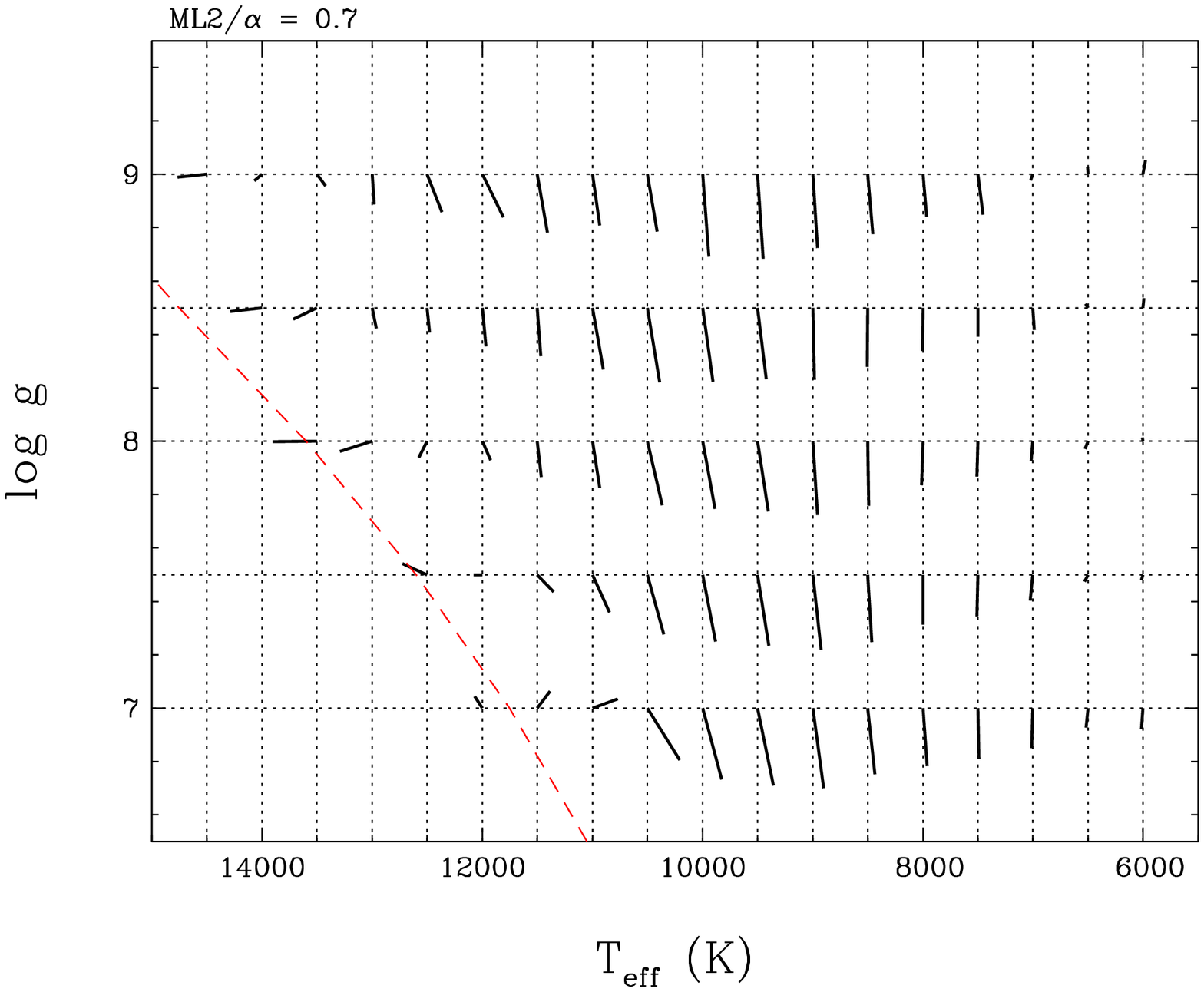}}
\caption{Similar to Fig.~\ref{fg:f_shifts} but with a reference grid of 1D
  spectra relying on the ML2/$\alpha$ = 0.6 (top) and 0.7 (bottom)
  parameterisation of the MLT. Tabulated values are available in the online
  Appendix B.
\label{fg:f_shifts_ML16-ML17}}
\end{center}
\end{figure}

In Fig.~\ref{fg:f_shifts_ML16-ML17}, we present 3D atmospheric parameter
corrections with reference 1D grids relying on the ML2/$\alpha$ = 0.6 and 0.7
parameterisation. The $\log g$ corrections are similar in strength although
the $T_{\rm eff}$ corrections are shifted. The ML2/$\alpha$ = 0.8 version
provides the best match between $\langle {\rm 3D}\rangle$ and 1D spectra at
the hot end of the sequence, although this result is sensitive to the
parameterisation of Eq.~(4). Nevertheless, the ML2/$\alpha$ = 0.6 definition
renders an inferior match to the 3D results that can not be improved much by
changing the line core removal procedure. At lower temperatures, 3D $T_{\rm
  eff}$ values are best matched with 1D models of lower convective efficiency,
hence it appears that the MLT calibration may be a function of $T_{\rm eff}$
and $\log g$. In particular, in the range of the ZZ Ceti instability strip at
$T_{\rm eff} \sim 12,000$~K, the ML2/$\alpha$ = 0.7 version provides $T_{\rm
  eff}$ values that are closer to the 3D results.

\begin{figure}[!h]
\begin{center}
\includegraphics[bb=38 142 572 642,width=3.2in]{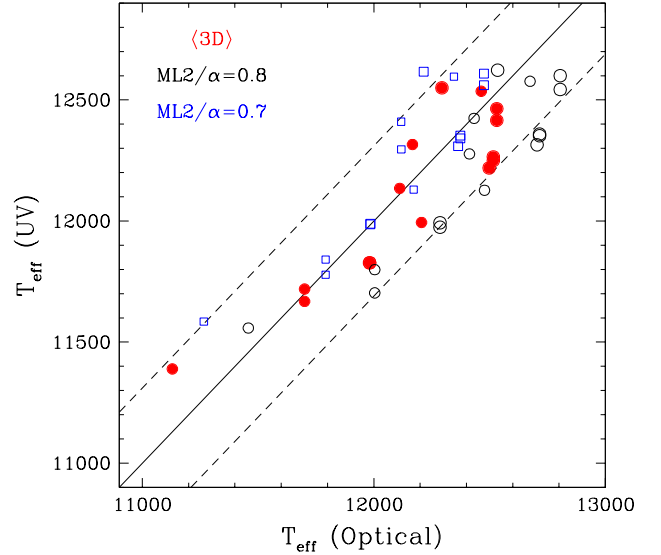}
\caption{Comparison of $T_{\rm eff}$ for ZZ Ceti stars derived from near-UV
  spectra and compared with the optical determinations. Optical log g values
  are assumed in the determination of the UV temperatures. The size of the
  symbols reflects the different weights assigned to the UV spectra. The solid
  line represents a perfect match between UV and optical temperatures, while
  the dashed lines represent the $\sim$350~K uncertainty allowed by the
  optical analysis. We rely on the $\langle {\rm 3D}\rangle$ (red filled
  circles), 1D ML2/$\alpha$ = 0.8 (black open circles), and 1D ML2/$\alpha$ =
  0.7 spectra (blue open squares). This is an updated version of Fig.~8 from
  \citet{TB10}.
\label{fg:f_UV}}
\end{center}
\end{figure}

The MLT calibration for white dwarf model atmospheres has been historically
performed from the comparison of effective temperatures measured independently
from optical and near-UV spectra \citep{bergeron95}. The ML2/$\alpha$ = 0.6
parameterisation proposed by \citet{bergeron95} has been widely employed over
the years. \citet{TB10} relied on the same technique to recalibrate the MLT to
ML2/$\alpha$ = 0.8 in light of their new model atmospheres with the
\citet{TB09} Stark broadening calculations. The authors also found that the
ML2/$\alpha$ $\sim$ 0.8 parameterisation produces the smoothest $\log g$
distribution as a function of $T_{\rm eff}$, i.e. without unexpected gaps or
clumps of stars in observed samples, which is a further internal consistency
check of the calibration.

We have updated the comparison of near-UV and optical parameters by using new
optical spectra \citep{gianninas11}. The set of UV spectra comes from the
Hubble Space Telescope (HST) and International Ultraviolet Explorer (IUE)
near-UV observations \citep{holberg03}. We first fit the Balmer lines in the
optical spectra to find $T_{\rm eff}$ (optical) and $\log g$ values. The
atmospheric parameters cannot be fitted simultaneously in the near-UV
continuum, hence we fix $\log g$ to the best-fit value in the optical, and
then determine $T_{\rm eff}$ (UV) from the near-UV
continuum. Fig.~\ref{fg:f_UV} demonstrates first of all that by relying on the
most recent optical observations, the optimal ML2/$\alpha$ value appears to
lie somewhere between 0.7 and 0.8 in this $T_{\rm eff}$ range. In the case of
the $\langle {\rm 3D}\rangle$ spectra, the comparison of the optical and
near-UV $T_{\rm eff}$ is not intended to calibrate the models but rather as an
internal check of the accuracy of the 3D structures. It is very interesting to
observe that $\langle {\rm 3D}\rangle$ spectra have consistent $T_{\rm eff}$
values in different wavelength regions. Since the near-UV flux at $T_{\rm eff}
\sim 12,000$~K is formed relatively deep in the photosphere ($\tau_{\rm R}
\sim$ 1) compared to the Balmer lines, it implies that the predicted structure
of the photosphere is consistent with the observations.

In light of the above results, we must find a way to combine the grids of
$\langle {\rm 3D}\rangle$ and 1D spectra. We observe that for ML2/$\alpha$ =
0.7 and 0.8 (Figs.~\ref{fg:f_shifts} and \ref{fg:f_shifts_ML16-ML17}), both
$\log g$ and $T_{\rm eff}$ corrections are fairly small at the hot end of the
3D sequence, which would appear to confirm that our choice of a boundary is
adequate. Since 3D $T_{\rm eff}$ corrections are sensitive to the line core
removal procedure, we must be cautious about the significance of this
agreement. We have chosen to make the connection at the $T_{\rm eff}$ which is
approximately the closest to the maximum strength of the H$\beta$ line (red
dashed line on Figs.~\ref{fg:f_shifts} and \ref{fg:f_shifts_ML16-ML17} for 1D
models). This is already a critical point in terms of line fitting since a
small discrepancy in the predicted and observed maximum strength of Balmer
lines ($\sim$1\% in the equivalent width) causes gaps or clumps of stars in
observed $\log g$ vs. $T_{\rm eff}$ distributions \citep{TB11}. Furthermore,
the $\chi^2$ fitting procedure often provides two valid solutions on both the
hot and cool side of the maximum, hence it is difficult in any case to obtain
precise atmospheric parameters in this region.  The combination of the
$\langle {\rm 3D}\rangle$ and 1D grids at this position allows to obtain
smooth $\log g$ vs. $T_{\rm eff}$ distributions outside the transition regime.

\subsection{3D atmospheric parameter correction functions}

We have derived functions that can be used to convert spectroscopically
determined 1D atmospheric parameters to their 3D counterparts. The recursive
fitting procedure is detailed in \citet{allende13}. Fortran 77 and IDL
functions are available in the online Appendices C and D, respectively. One can
also rely on the data from Figs.~\ref{fg:f_shifts} and
\ref{fg:f_shifts_ML16-ML17} given in the online Appendix B to derive
alternative functions. Our independent variables are defined as

\begin{equation}
g_{\rm X} = \log g~{\rm [cgs]} - 8.0 ~,
\end{equation}

\begin{equation}
T_{\rm X} = \frac{T_{\rm eff}~{\rm [K]} - 10,000}{1000} ~,
\end{equation}

{\noindent}and the fitting functions for the ML2/$\alpha$ = 0.7 (Eq.~(7) and (8))
and 0.8 (Eq.~(9) and (10)) parameterisation of the MLT are given below with the
numerical coefficients identified in Table 1.

\begin{dmath}
\frac{\Delta T_{\rm eff}}{1000} {\rm ~[ML2/}\alpha~{\rm = 0.7]} = a_0+[a_1+a_4T_{\rm
  X}+(a_5+a_6T_{\rm X}+a_7g_{\rm X})g_{\rm X}]\exp[-a_2(T_{\rm X}-a_3)^2] ~,
\end{dmath}

\begin{dmath}
\Delta \log g {\rm ~[ML2/}\alpha~{\rm = 0.7]} =
b_0+b_1\exp\left\{-\big(b_2+(b_4+b_6\exp[-b_7(T_{\rm X}-b_8)^2])T_{\rm
  X}+(b_5+b_9T_{\rm X}+b_{10}g_{\rm X})g_{\rm X}\big)^2(T_{\rm
  X}-b_3)^2\right\} ~.
\end{dmath}

\begin{dmath}
\frac{\Delta T_{\rm eff}}{1000} {\rm ~[ML2/}\alpha~{\rm = 0.8]} =
c_0+(c_1+c_6T_{\rm X}+c_7g_{\rm X})\exp[-(c_2+c_4T_{\rm X}+c_5g_{\rm
    X})^2(T_{\rm X}-c_3)^2] ~,
\end{dmath}

\begin{dmath}
\Delta \log g {\rm ~[ML2/}\alpha~{\rm = 0.8]} = \big(d_0+d_4\exp[-d_5(T_{\rm
  X}-d_6)^2]\big)+d_1\exp\left\{-d_2\big[T_{\rm X}-(d_3+d_7\exp(-[d_8+d_{10}T_{\rm
  X}+d_{11}g_{\rm X}]^2[T_{\rm X}-d_9]^2))\big]^2\right\} ~.
\end{dmath}

 \begin{table}[h!]
 \caption{Coefficients for fitting functions}
 \label{tab1}
 \begin{center}
 \begin{tabular}{llll}
\hline
\hline
Coeff. & $\Delta T_{\rm eff}$ [ML2/$\alpha$ = 0.7] & Coeff. & $\Delta\log g$ [ML2/$\alpha$ = 0.7] \\
\hline
$a_{0}$ & $-$1.0461690E$-$03 & $b_{0}$ &  1.1922481E$-$03 \\
$a_{1}$ & $-$2.6846737E$-$01 & $b_{1}$ & $-$2.7230889E$-$01 \\
$a_{2}$ &  3.0654611E$-$01 & $b_{2}$ & $-$6.7437328E$-$02 \\
$a_{3}$ &  1.8025848E+00 & $b_{3}$ & $-$8.7753624E$-$01 \\
$a_{4}$ &  1.5006909E$-$01 & $b_{4}$ &  1.4936511E$-$01 \\
$a_{5}$ &  1.0125295E$-$01 & $b_{5}$ & $-$1.9749393E$-$01 \\
$a_{6}$ & $-$5.2933335E$-$02 & $b_{6}$ &  4.1687626E$-$01 \\
$a_{7}$ & $-$1.3414353E$-$01 & $b_{7}$ &  3.8195432E$-$01 \\
...    & ...            & $b_{8}$ & $-$1.4141054E$-$01 \\
...    & ...            & $b_{9}$ & $-$2.9439950E$-$02 \\
...    & ...            & $b_{10}$ & 1.1908339E$-$01 \\
\hline
Coeff. & $\Delta T_{\rm eff}$ [ML2/$\alpha$ = 0.8] & Coeff. & $\Delta\log g$ [ML2/$\alpha$ = 0.8] \\
\hline
$c_{0}$ &  1.0947335E$-$03 & $d_{0}$ &  7.5209868E$-$04 \\
$c_{1}$ & $-$1.8716231E$-$01 & $d_{1}$ & $-$9.2086619E$-$01 \\
$c_{2}$ &  1.9350009E$-$02 & $d_{2}$ &  3.1253746E$-$01 \\
$c_{3}$ &  6.4821613E$-$01 & $d_{3}$ & $-$1.0348176E+01 \\
$c_{4}$ & $-$2.2863187E$-$01 & $d_{4}$ &  6.5854716E$-$01 \\
$c_{5}$ &  5.8699232E$-$01 & $d_{5}$ &  4.2849862E$-$01 \\
$c_{6}$ & $-$1.0729871E$-$01 & $d_{6}$ & $-$8.8982873E$-$02 \\
$c_{7}$ &  1.1009070E$-$01 & $d_{7}$ &  1.0199718E+01 \\
...    & ...            & $d_{8}$ &  4.9277883E$-$02 \\
...    & ...            & $d_{9}$ & $-$8.6543477E$-$01 \\
...    & ...            & $d_{10}$ & 3.6232756E$-$03 \\
...    & ...            & $d_{11}$ &$-$5.8729354E$-$02 \\
\hline
\end{tabular} 
\end{center} 
\end{table}

Our fitting functions provide negligible 3D corrections for DA white dwarfs
with $T_{\rm eff}$ values outside the range of our 3D simulations. We are also
aware that our 3D grid does not extend to high and low enough gravities to
cover some of the most extreme and interesting white dwarf observations
\citep[see, e.g.,][]{kepler07,hermes13}. However, our fitting functions should
be able to provide reasonable estimates since 3D parameter corrections are, in
most cases, fairly regular as a function of $\log g$, except for the
transition region between convective and radiative atmospheres.

We do not provide a function for ML2/$\alpha$ = 0.6 since the connection of
our $\langle {\rm 3D}\rangle$ spectra with hotter 1D spectra is
problematic. We suggest instead that 1D models with ML2/$\alpha$ = 0.7 or 0.8
should be used. Nevertheless, we provide tabulated values of the 3D
corrections for ML2/$\alpha$ = 0.6 in the Appendix B.

\begin{figure}[!h]
\begin{center}
\includegraphics[bb=38 172 572 642,width=3.2in]{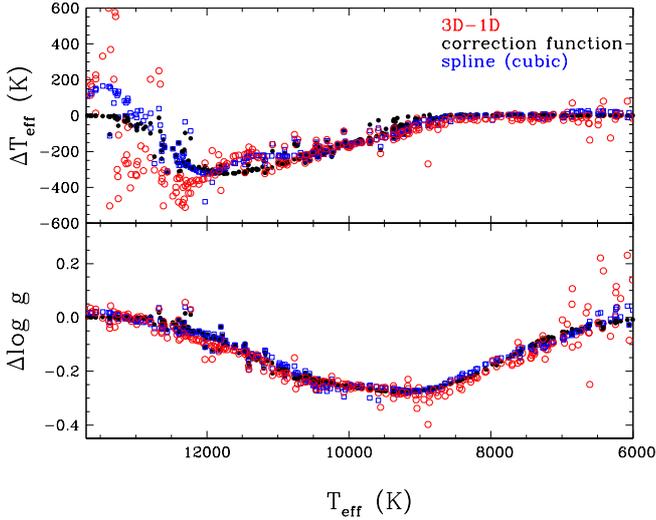}
\caption{3D vs. 1D $T_{\rm eff}$ (top) and $\log g$ (bottom) from actual 3D
  and 1D fits (red open circles) and by using our correction function (Eq.~(9)
  and (10)) to modify the 1D parameters (black filled circles). We also employ
  a cubic spline interpolation of the 3D corrections (blue open squares) to
  modify the 1D parameters.
\label{fg:f_compar}}
\end{center}
\end{figure}

In Fig.~\ref{fg:f_compar}, we compare $T_{\rm eff}$ and $\log g$ values for
the \citet{gianninas11} sample (see Sect.~4.1) found by fitting directly the
observations with our $\langle {\rm 3D}\rangle$ and 1D spectra and from
converting the 1D parameters using our fitting functions (Eq.~(9) and (10)). As a
reference, we also convert the 1D parameters by employing a cubic spline
interpolation with the data from Fig.~\ref{fg:f_shifts}. This method preserves
more details compared to our fitting functions. We note that additional
sequences at $\log g =$ 7.75 and 8.25 are part of the 1D grid and that line
cores are not removed from the 1D fits. Nevertheless, it is shown that the
atmospheric parameters are fairly similar in all cases, and that our fitting
functions are adequate to provide precise 3D parameters. We have chosen
fitting functions that do not preserve all details of the 3D $T_{\rm eff}$
corrections unlike, for instance, the spline interpolation. This choice was
made because the small scale fluctuations are below the model (e.g. line core
removal procedure) and observational uncertainties, and it would be difficult to
justify that they are real physical features.

At the hot end of the sequence, there are some discrepancies in the $T_{\rm
  eff}$ values found from fits with $\langle {\rm 3D}\rangle$ spectra and
tabulated corrections (with either the spline interpolation or fitting
functions).  We believe that the 3D corrections drawn from
Fig.~\ref{fg:f_shifts} smooth out some of the issues obtained with the
monochromatic flux interpolation in the predicted spectra. The latter is
difficult to perform in this range of $T_{\rm eff}$ because the wavelength
regions close to the line cores are predicted differently in the 1D and
$\langle {\rm 3D}\rangle$ spectra. At the cool end of the sequence, it appears
that the line core removal procedure enhances the scatter in the $\langle {\rm
  3D}\rangle$ fits.  In light of these results, we have decided to rely on the
correction functions (Eq.~(9) and (10)) for the rest of this work.

\section{Astrophysical implications}

In order to study the astrophysical implications of our improved predicted
spectra for convective DA white dwarfs, we have chosen to review the
properties of two well studied samples, that of the White Dwarf Catalog
\citep{ms99} and the Sloan Digital Sky Survey \citep[][hereafter
  SDSS]{sdss}. In particular, we base our study on the spectroscopic
analyses of \citet[][hereafter GB11]{gianninas11} for the White Dwarf Catalog
and \citet[][hereafter TB11]{TB11} for the SDSS, where the sample is
  derived from the Data Release 4. The advantage of these studies is that
they rely on the same 1D model atmospheres (ML2/$\alpha$ = 0.8) and the same
fitting method as we use in this work. Furthermore, in both cases the authors
made a thoughtful examination of the data in order to identify subtypes, such
as binaries and magnetic white dwarfs. This allows for a better precision in
the determination of the mean properties of the samples. While both surveys
have been enlarged with newly observed DA white dwarfs \citep[][and
  A. Gianninas, private communication]{kleinman13}, we believe that our
selected samples are large and complete enough to study the 3D effects on the
predicted atmospheric parameters.

We convert $\log g$ values into stellar masses using the evolutionary models
with thick hydrogen layers of \citet{fontaine01} below $T_{\rm eff} <$ 30,000
K and of \citet{wood95} above this temperature.  Low-mass white dwarfs, below
0.46 $M{_\odot}$ and $T_{\rm eff} <$ 50,000~K, are likely helium core white
dwarfs, and we rely instead on evolutionary models from \citet{althaus01}. For
masses higher than 1.3 $M{_\odot}$, we use the zero temperature calculations
of \citet{hamada}.

\subsection{White Dwarf Catalog sample}

We review the mass distribution of the \citet{gianninas11} sample of 1265
bright DA white dwarfs ($V \lesssim 17.5$) drawn from the online version of
the McCook \& Sion White Dwarf
Catalog\footnote{http://www.astronomy.villanova.edu/WDCatalog/} (hereafter
WDC). High S/N observations ($\langle{\rm S/N}\rangle$ $\sim$ 70) were secured
for all objects in the sample over a period of $\sim$20 years at different
sites (see Sect.~2 of GB11). Therefore, observations in the GB11 sample do not
have a homogeneous instrumental setup, with a resolution varying from $\sim$3
to 12 \AA~(FWHM).  Our starting point are the objects identified in Table~5 of
GB11. We rely on the same observations and the same grid of 1D model
atmospheres to determine the atmospheric parameters. We took their solutions
for DAO and hot DA white dwarfs with the Balmer-line problem \citep{werner96}
where they employed NLTE models with carbon, nitrogen and oxygen
\citep{gianninas10}. The addition of these elements has a cooling effect on
the upper layers and the predicted line cores are in better agreement with the
observations. Therefore, our analysis only differs from that of GB11
considering that we also derive 3D atmospheric parameters.

\begin{figure}[!h]
\begin{center}
\includegraphics[bb=38 172 572 642,width=3.2in]{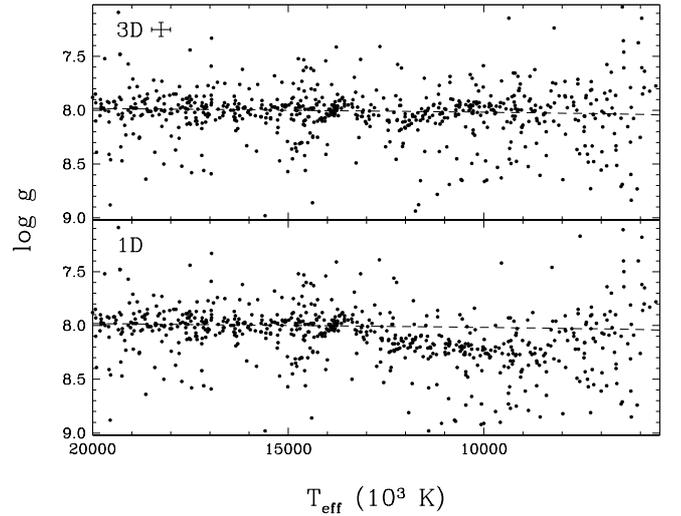}
\caption{$\log g$ versus $T_{\rm eff}$ distribution for DA white dwarfs in the
  GB11 sample derived from $\langle$3D$\rangle$ (top) and 1D spectra
  (bottom). An evolutionary model from \citet{fontaine01} at the median mass
  of the sample (0.61 $M_{\odot}$, $13,000 < T_{\rm eff}$ (K) $< 40,000$) is
  shown as a dashed line. The error bars on the top left represent the mean
  uncertainties in this $T_{\rm eff}$ range.
\label{fg:f_logg_WDC}}
\end{center}
\end{figure}

\begin{figure}[!h]
\begin{center}
\includegraphics[bb=38 172 572 642,width=3.2in]{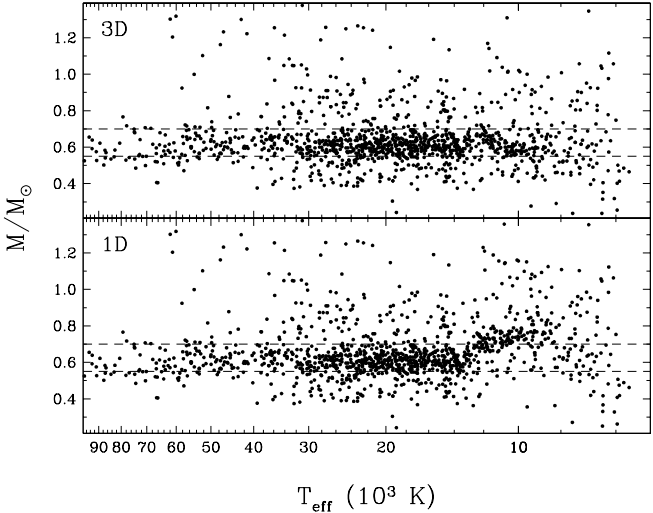}
\caption{Mass versus $T_{\rm eff}$ (logarithm scale) distribution for DA white
  dwarfs in the GB11 sample derived from $\langle$3D$\rangle$ (top) and 1D
  spectra (bottom). Lines of constant mass at 0.55 $M{_\odot}$ and 0.7
  $M{_\odot}$ are shown as a reference.
\label{fg:f_mass_WDC}}
\end{center}
\end{figure}

We present in Fig.~\ref{fg:f_logg_WDC} and \ref{fg:f_mass_WDC} the GB11 sample
$\log g$ and mass distributions as a function of $T_{\rm eff}$ employing
either 1D or 3D model atmospheres. The distributions are identical for
radiative atmospheres ($T_{\rm eff} \gtrsim 15,000$ K). The distributions
relying on $\langle {\rm 3D}\rangle$ spectra are much more stable as a
function of $T_{\rm eff}$ and the high-$\log g$ problem is to a very large
extent removed. Nevertheless, masses at $T_{\rm eff} \sim 12,000$~K appear
slightly larger than the sample average. We review this issue more
  quantitatively in Sect.~5.

\begin{figure}[!]
\captionsetup[subfigure]{labelformat=empty}
\begin{center}
\subfloat[]{
\includegraphics[bb=38 172 572 592,width=3.2in]{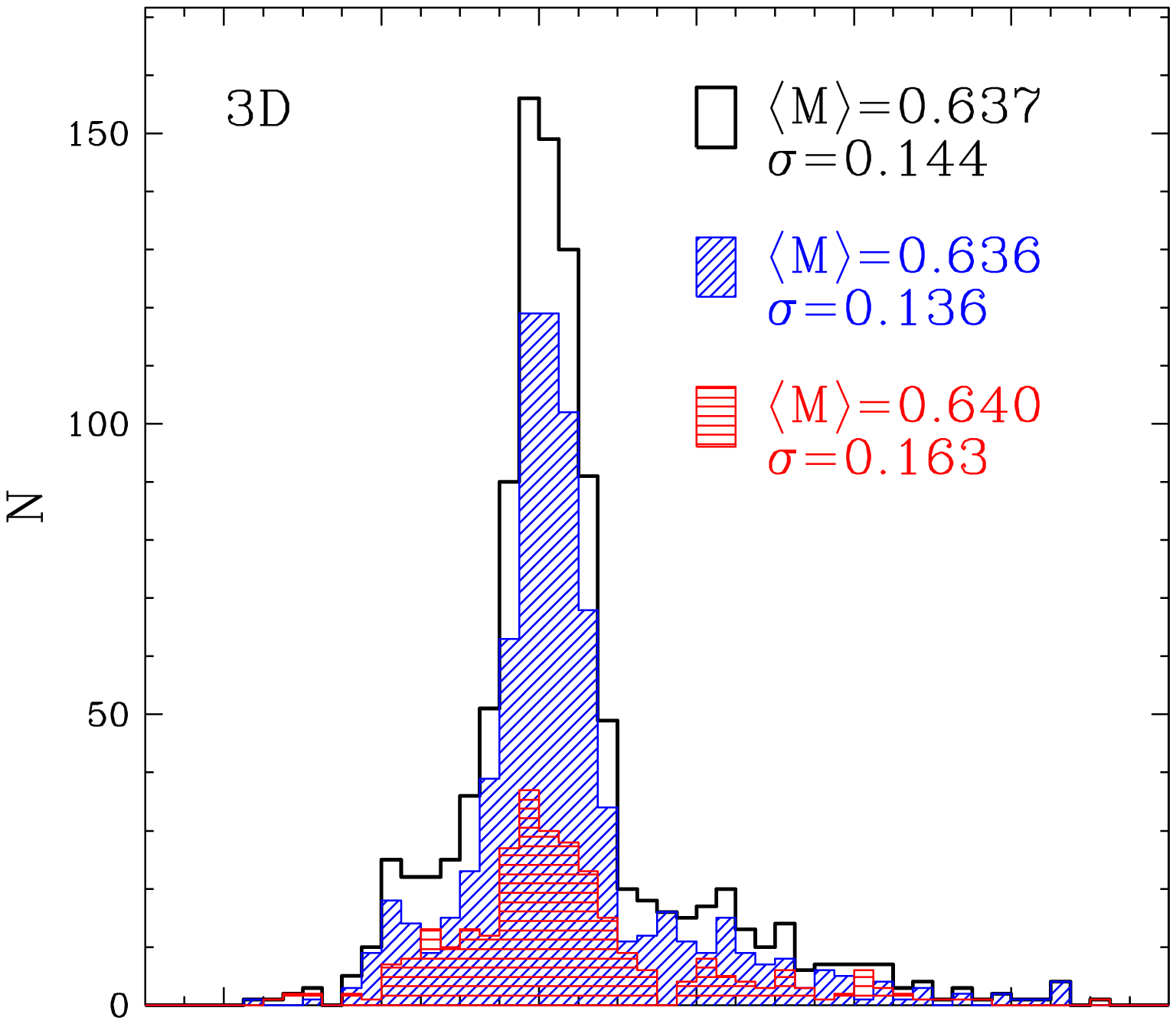}}\\
\subfloat[]{
\includegraphics[bb=38 172 572 502,width=3.2in]{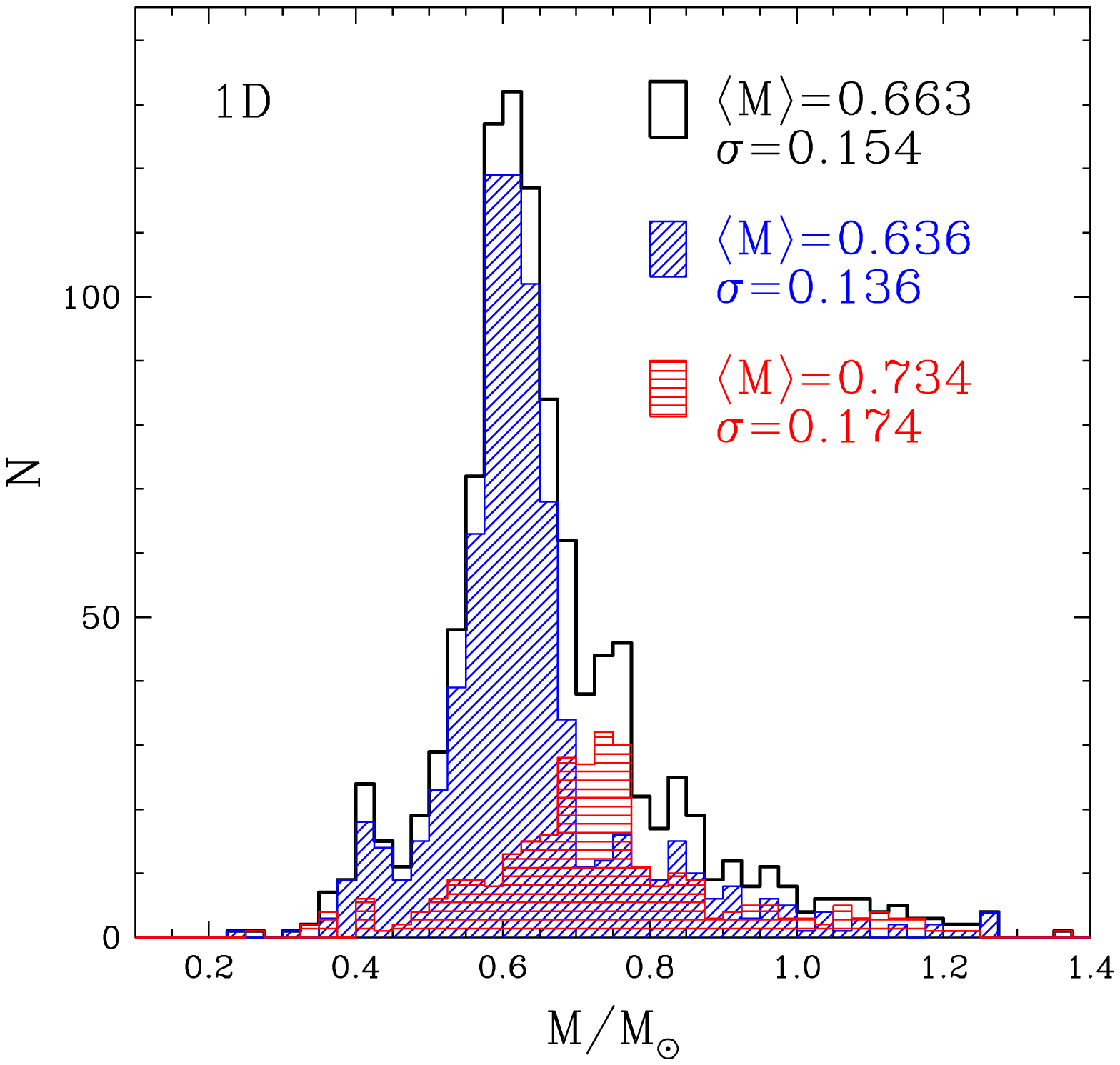}}
\caption{{\it Top:} Mass histogram for DA stars in the GB11 sample with
  $T_{\rm eff} < $ 40,000~K (black empty histogram) from $\langle$3D$\rangle$
  spectra. We also show the sub-distributions for $13,000 < T_{\rm eff}$ (K)
  $< 40,000$ (blue histogram, dashed diagonally, radiative atmospheres) and $T_{\rm eff} <
  13,000$~K (red histogram, dashed horizontally, convective atmospheres). The number of stars is
  given on the left-hand scale. The mean masses and standard deviations are
  indicated in the panels in units of solar masses. Binaries and magnetic
  objects were removed from the distributions. {\it Bottom:} Same as above but
  with 1D spectra.
\label{fg:f_hist_WDC}}
\end{center}
\end{figure}

The mass histograms for the GB11 sample using both 1D and 3D models are shown
in Fig.~\ref{fg:f_hist_WDC}. We have restricted the analysis to objects below
$T_{\rm eff} < 40,000$~K for an easier comparison with the SDSS sample even
though hot white dwarfs in the GB11 sample have accurate atmospheric
parameters. Furthermore, we have removed all objects identified as binaries,
including DA+dM composite spectra and double degenerates, and objects with
evidences of Zeeman splitting. These subtypes typically have larger
uncertainties on their atmospheric parameters and appear to have different
mass distributions than typical single DAs, either due an observational bias
or intrinsic properties (\citealt{kawka07}; TB11). We have used a threshold at
13,000~K to separate between the regime of radiative (blue colour in
Fig.~\ref{fg:f_hist_WDC}) and convective (red) atmospheres. The 1D mean masses
for convective and radiative white dwarfs have a significant offset of $\Delta
M/M{_\odot}$ = +0.10, which is largely corrected by the 3D models, where the
offset is not significant.  The properties of the standard deviation are
discussed in Sect.~5.

\subsection{Sloan Digital Sky Survey}

We rely on the sample of \citet{E06} drawn from the Data Release 4 of the
SDSS, which has been reviewed in numerous studies \citep[][and
  TB11]{kepler07,degennaro08}. In particular, our starting point are the 3072
DA stars in Table~1 of TB11, who improved upon previous analyses by including
the Stark broadening profiles of \citet{TB09} and by performing a careful
visual inspection to identify subtypes and calibration problems. In the
following, we identify this sample as E06/TB11. We note that the
\citet{kleinman13} sample drawn from the DR7 approximately doubles the size of
the previous \citet{E06} sample. The \citet{kleinman13} analysis relies on
different 1D model atmospheres, fitting procedures and manual inspection
methods to identify subtypes. While their study should be equivalent to that
of TB11, we would need first of all to compare both approaches, which is out
of the scope of this work.

White dwarf spectra in the sample of \citet{E06} were identified from a
combination of the object $ugriz$ colours, the absence of a redshift, and an
automated comparison of the photometric and spectroscopic data with white
dwarf models. These steps are quite robust in recovering most single DA white
dwarfs with $T_{\rm eff} \gtrsim 8000$~K observed spectroscopically by the
SDSS. However, the completeness of the SDSS spectroscopic survey itself can be
anywhere between 15\% and 50\%, in part due to the neglect of blended point
sources, as well as an incomplete spectroscopic follow-up of the point sources
identified in the SDSS fields \citep{E06,degennaro08,kleinman13}.
 
TB11 found that a cutoff at S/N $> 15$, which we also apply here, allowed for
the best compromise between the size of the sample and the precision of its
mean properties. The sole difference in comparison to the TB11 analysis is
that we rely on the most recent DR9 data reduction, instead of the one from
DR7, for both the spectroscopic and photometric data obtained from the SDSS
DR9 Science Archive Server and Sky Server\footnote{dr9.sdss3.org,
  skyserver.sdss3.org/dr9}. The spectra, which cover a wavelength range of
3800 to $\sim$10,000~\AA~with a resolution of $R\sim 1800$, appeared to have
some calibration issues up to the DR7 (\citealt{kleinman04}; \citealt{E06};
TB11; GB11). For a sample of 89 stars observed by both the SDSS and GB11, the
Fig.~17 of TB11 highlights a systematic $\log g$ difference between both data
sets. Since the models and the fitting procedure are identical, differences
must lie in the observations. This issue is mostly seen for hot white dwarfs
and given the available objects in common, the samples are roughly in
agreement within the uncertainties for cool convective objects. Nevertheless,
we feel that it is especially important to rely on the latest data
reduction\footnote{For 42 DAO and hot white dwarfs ($T_{\rm eff} >$ 40,000~K)
  with the Balmer line problem and S/N $>$ 15, we use the parameters from TB11
  with the DR7 data.}.

\begin{figure}[!h]
\begin{center}
\includegraphics[bb=38 172 572 642,width=3.2in]{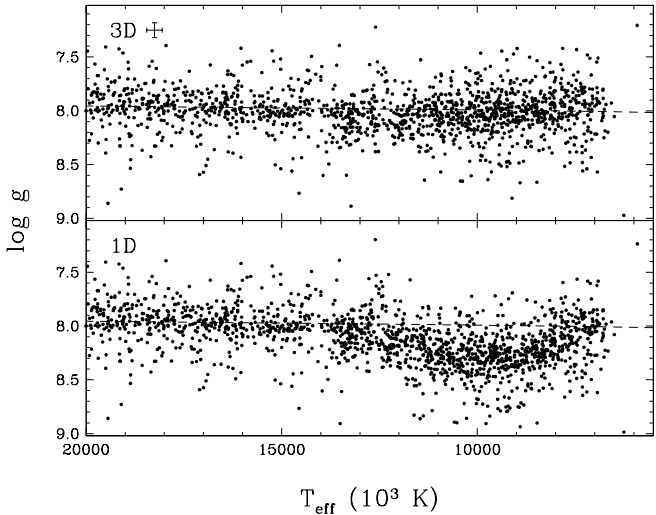}
\caption{$\log g$ versus $T_{\rm eff}$ distribution for DA white dwarfs with
  S/N $>$ 15 in the SDSS E06/TB11 sample derived from $\langle$3D$\rangle$
  (top) and 1D (bottom) spectra. An evolutionary model from \citet{fontaine01}
  at the median mass of the sample (0.59 $M_{\odot}$, $13,000 < T_{\rm eff}$
  (K) $< 40,000$) is shown as a dashed line. The error bars on the top left
  represent the mean uncertainties in this $T_{\rm eff}$ range.
\label{fg:f_logg_SDSS}}
\end{center}
\end{figure}

\begin{figure}[!h]
\begin{center}
\includegraphics[bb=38 172 572 642,width=3.2in]{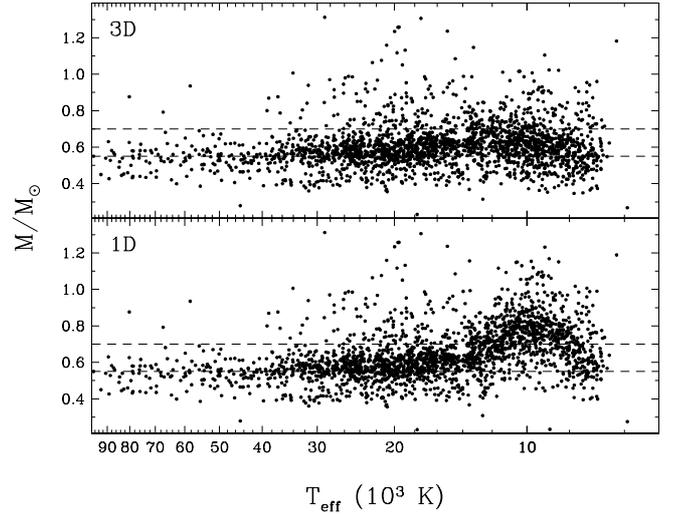}
\caption{Mass versus $T_{\rm eff}$ distribution for DA white
  dwarfs in the SDSS E06/TB11 sample derived from $\langle$3D$\rangle$ (top) and 1D
  spectra (bottom). Lines of constant mass at 0.55 $M{_\odot}$ and 0.7
  $M{_\odot}$ are shown as a reference.
\label{fg:f_mass_SDSS}}
\end{center}
\end{figure}

In Fig.~\ref{fg:f_logg_SDSS} and \ref{fg:f_mass_SDSS} we present the $\log g$
and mass distributions as a function of $T_{\rm eff}$ for the SDSS E06/TB11
sample (S/N $> 15$) with both the 1D and $\langle {\rm 3D}\rangle$
spectra. For cooler convective objects, the 1D $\log g$ distribution
illustrates the classic high-$\log g$ problem with a significant downturn. The
problem is largely removed from the distribution relying on 3D model
atmospheres. Nevertheless, the mass distribution still shows some
substructures with masses that appear too high at $T_{\rm eff} \sim 12,000$~K,
similarly to what is observed for the GB11 sample.

\begin{figure}[!]
\captionsetup[subfigure]{labelformat=empty}
\begin{center}
\subfloat[]{
\includegraphics[bb=38 172 572 592,width=3.2in]{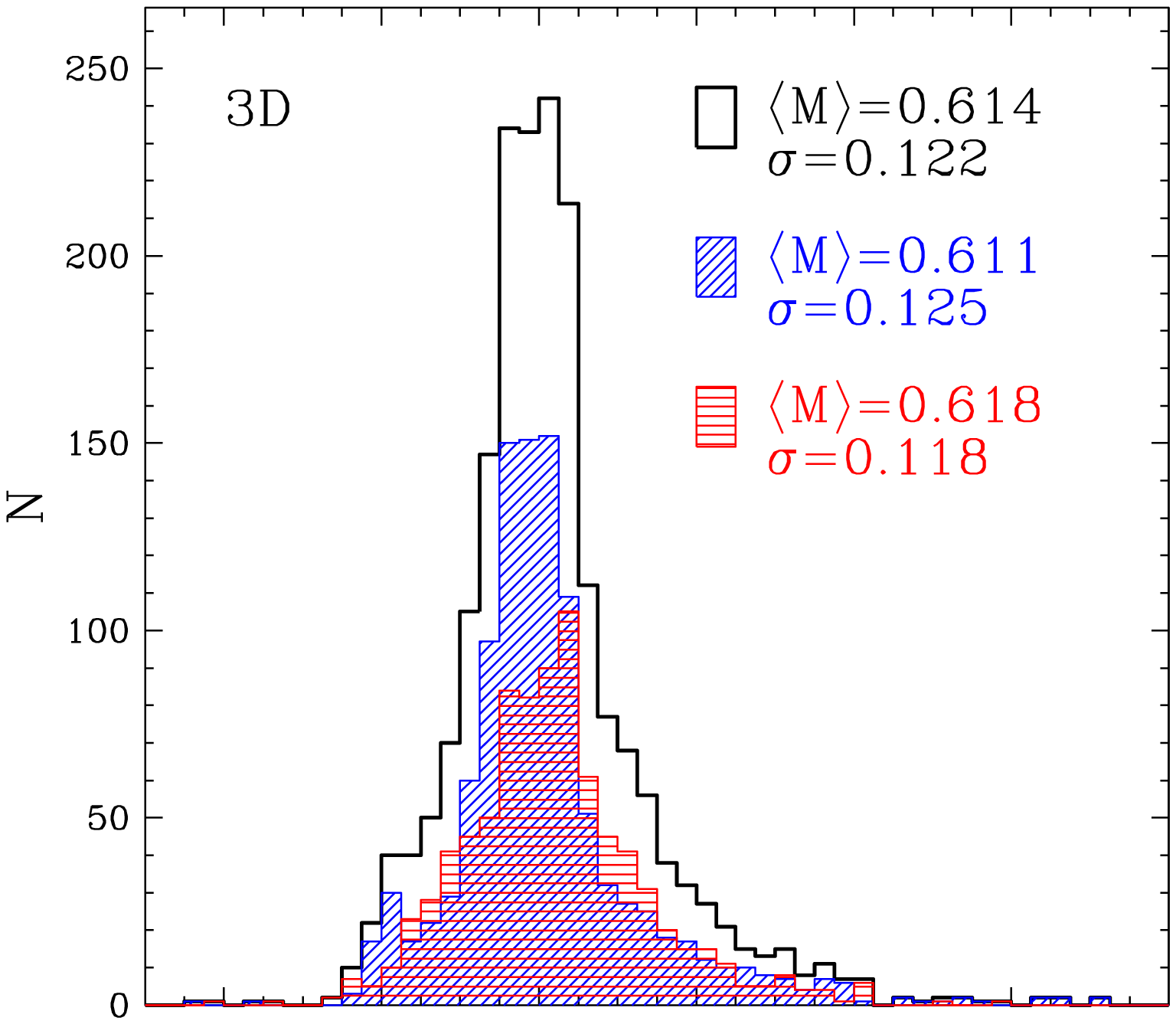}}\\
\subfloat[]{
\includegraphics[bb=38 172 572 502,width=3.2in]{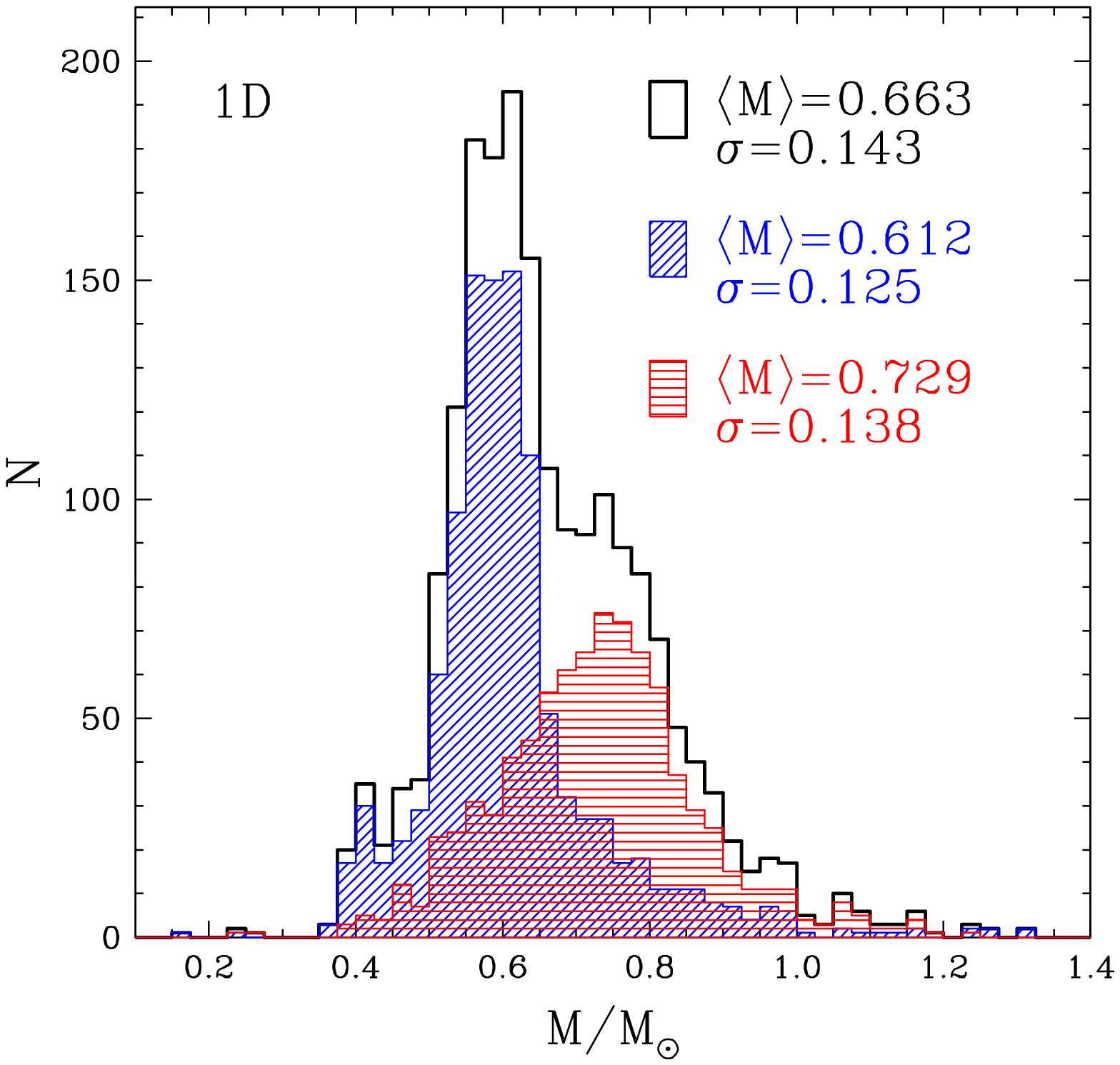}}
\caption{Similar to Fig.~\ref{fg:f_hist_WDC} but for the SDSS E06/TB11 sample
  with $T_{\rm eff} < $ 40,000~K and S/N $>$ 15 (black empty
    histograms). We also highlight the sub-distributions for $13,000 < T_{\rm
      eff}$ (K) $< 40,000$ (blue histograms, radiative atmospheres) and $T_{\rm
      eff} < 13,000$~K (red histograms, convective atmospheres). Binaries and
    magnetic objects were removed from the distributions.
\label{fg:f_histo_SDSS}}
\end{center}
\end{figure}

As in TB11, we restrict our determination of the mean mass to objects with
$T_{\rm eff} < 40,000$~K. This cutoff is applied in part because it is
difficult to identify metal-rich objects with the Balmer-line problem at the
low average S/N of the SDSS observations.  Furthermore, we remove all
identified binaries and magnetic objects. Fig.~\ref{fg:f_histo_SDSS} includes
mass histograms for both the hot radiative (blue, $T_{\rm eff} > 13,000$~K)
and cool convective (red, $T_{\rm eff} < 13,000$~K) atmospheres. The first
observation is that the mass distribution of hot radiative white dwarfs is
nearly identical to the one found in TB11 using the DR7 instead of the DR9
data reduction. The 1D mass distribution below $T_{\rm eff}$ = 13,000~K has a
significantly higher mean value, with $\Delta M/M{_\odot}$ = +0.12. In the
case of the 3D mass distributions, this shift decreases to a value of
+0.01. This outcome is very similar to that of the GB11 sample.

\subsection{Photometric samples}

Photometric observations of white dwarfs allow for the determination of
independent $T_{\rm eff}$ values \citep{koester79}. When trigonometric parallax measurements
are available, it is also possible to constrain $\log g$
\citep{bergeron01}. Since our $\langle {\rm 3D}\rangle$ spectra predict roughly
the same optical colours as the 1D models, earlier photometric studies are
unchanged and we can compare directly these results to our spectroscopic
parameters.

\begin{figure}[!h]
\begin{center}
\includegraphics[bb=38 172 572 642,width=3.2in]{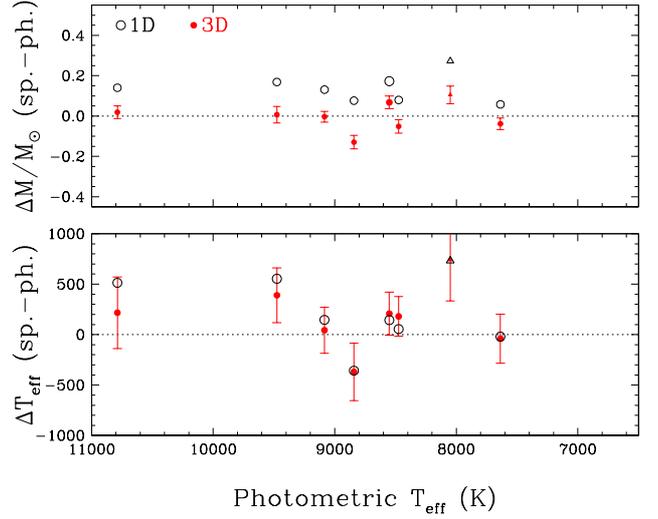}
\caption{Comparison of the photometric and spectroscopic masses (top) and
  $T_{\rm eff}$ (bottom) for DA white dwarfs in the \citet{subasavage09}
  sample. We rely on both $\langle$3D$\rangle$ (filled red circles and error
  bars) and 1D (open black circles) spectra. The error bars are omitted in the
  1D case for clarity. The object represented by a triangle (LHS 4040)
  possibly has some contamination from a nearby M dwarf. The horizontal dotted
  lines correspond to a perfect match between spectroscopic and photometric
  parameters. This is an updated version of Fig.~12 from \citet{TB10}.
\label{fg:f_Subasavage}}
\end{center}
\end{figure}

\citet{koester09} and \citet{TB10} have reviewed the current photometric
constraints on the atmospheric parameters of convective DAs with the
perspective of finding a solution to the high-$\log g$ problem. We can
therefore use their studies as a starting point. There is currently a lack of
parallax samples that are both precise and large. Therefore, \citet{TB10}
concluded that the parallax and $VRIJHK$ photometric sample of eight DA stars
from \citet{subasavage09} provided the most precise independent constraint on
the individual masses of cool DA stars. In Fig.~\ref{fg:f_Subasavage}, we
update the Fig.~12 of \citet{TB10} by presenting a comparison of the
photometric and spectroscopic parameters relying on both the 3D and 1D model
atmospheres. The masses found from parallaxes are now in much better agreement
with those found spectroscopically by using $\langle {\rm 3D}\rangle$
spectra. However, given the small size of the sample and possible differences
in the internal composition of those white dwarfs influencing the mass-radius
relation, it is difficult to conclude further on the accuracy of the 3D model
atmospheres.

The photometric data also offers an unique opportunity to independently
measure $T_{\rm eff}$ values. In the case of the \citet{subasavage09} sample,
Fig.~\ref{fg:f_Subasavage} shows that photometric and spectroscopic $T_{\rm
  eff}$ are in a slightly better agreement in 3D.  However, the mixing-length
parameter in the 1D models could be updated to provide a closer agreement with
the 3D values.

\begin{figure}[!h]
\begin{center}
\includegraphics[bb=38 172 572 642,width=3.2in]{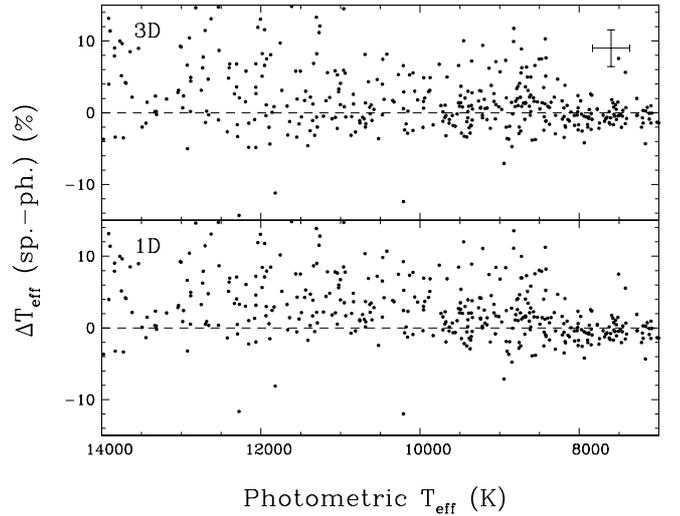}
\caption{Comparison of photometric and spectroscopic $T_{\rm eff}$ for DA
  white dwarfs in the SDSS E06/TB11 sample relying on $\langle$3D$\rangle$
  (top) and 1D (bottom) spectra. The mean uncertainty is shown with error
  bars on the top panel.
\label{fg:f_phot}}
\end{center}
\end{figure}

The $ugriz$ energy distribution is available for most SDSS objects in our
sample, and for convective objects, the predicted magnitudes are rather
sensitive to $T_{\rm eff}$, allowing for a precision on the photometric
$T_{\rm eff}$ of the order of 5$\%$.  Since parallaxes are not available, it
is not possible to individually constrain $\log g$ values.
Fig.~\ref{fg:f_phot} compares the spectroscopic and photometric $T_{\rm eff}$
for both the 1D and 3D model atmospheres. To prevent uncertainties due to
reddening, we only include objects with $g < 18$. Fig.~\ref{fg:f_phot}
demonstrates that 3D spectroscopic $T_{\rm eff}$ are in reasonable agreement
with the photometric values.  Since both the spectroscopic and photometric
SDSS data may be impacted by calibration issues, it is difficult to conclude
on the accuracy of 3D $T_{\rm eff}$ corrections from this
comparison. Nevertheless, the results of this section, along with the
comparison of optical and near-UV $T_{\rm eff}$ in Sect. 3.3, illustrate that
the corrections are certainly acceptable. A dedicated survey of high-precision
photometry and parallaxes for cool DA white dwarfs would help in further
understanding the accuracy of the model atmospheres.

\section{Discussion}

We have described in this work the effects of 3D model atmospheres on the
determination of the atmospheric parameters of DA white dwarfs. Our proposed
3D $\log g$ corrections are in agreement with those found in our previous
study limited to $\log g = 8$ (Paper II). We have shown that the mass
distribution of radiative and convective atmosphere white dwarfs are now in
quantitative agreement.  We have confirmed the results of earlier studies that
1D model atmospheres relying on the MLT are responsible for the high-$\log g$
problem and that 3D model atmospheres do not have this shortcoming. It
demonstrates clearly that one should rely on 3D model atmospheres, or 1D model
atmospheres with 3D correction functions, to have a precision higher than
$\sim$10-20\% on the mass and age of convective DA stars. In this section, we
attempt to further characterise the accuracy of our current grid of
$\langle$3D$\rangle$ spectra by comparing the results found for the SDSS
E06/TB11 and \citet{gianninas11} samples.

\subsection{Accuracy of the $\langle$3D$\rangle$ spectra}

It is not straightforward to compare the SDSS E06/TB11 and GB11 samples in
terms of the absolute value of the mean mass and the related standard
deviation. In addition to possible issues with the data calibration, the
samples have significantly different selection criteria and completeness
levels (\citealt{E06}; GB11). The SDSS sample is mostly magnitude-limited
while stars identified in the White Dwarf Catalog are drawn from various
surveys, including X-ray surveys like the one conducted with ROSAT which is
not magnitude limited. Since high-mass white dwarfs are fainter, it is perhaps
not surprising that the White Dwarf Catalog sample includes more massive
objects than the SDSS, possibly explaining the higher mean mass. The mass
dispersion is also significantly higher for the GB11 sample, perhaps also due
to the presence of a larger number of high-mass objects, but it is suspected
that the different instrumental setup and data reduction have an impact as
well.

\begin{figure}[!h]
\begin{center}
\includegraphics[bb=38 92 572 692,width=3.2in]{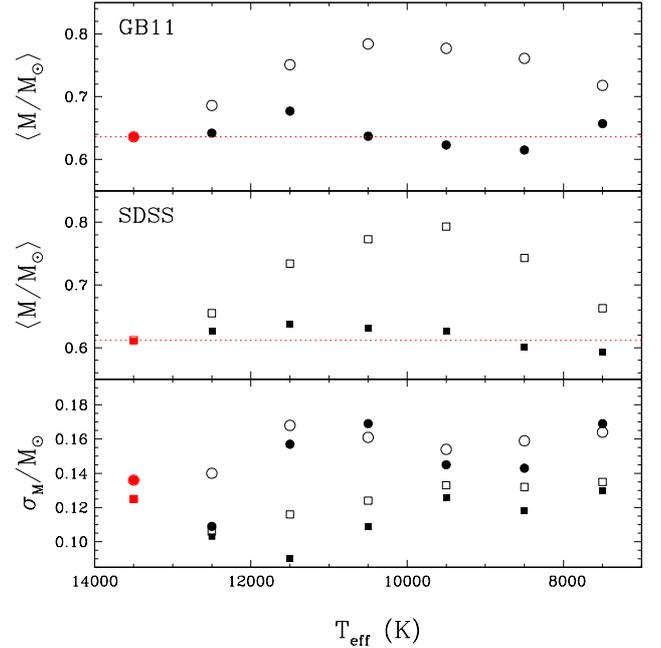}
\caption{Mean mass (top panels) and standard deviation (bottom panel) computed
  in 1000~K temperature bins for the GB11 (circles, top panel) and SDSS E06/TB11
  (squares, middle panel) samples. Values are derived from both
  $\langle$3D$\rangle$ (filled circles) and 1D spectra (open circles). The red
  circles on the left and the dotted lines represent the values for hotter
  radiative objects ($13,000 < T_{\rm eff}$ (K) $< 40,000$).
\label{fg:f_bin}}
\end{center}
\end{figure}

In Fig.~\ref{fg:f_bin}, we compare the mass distributions of the SDSS E06/TB11
and GB11 samples (Figs.~\ref{fg:f_mass_WDC} and \ref{fg:f_mass_SDSS}) in bins
of 1000~K to better understand mean mass fluctuations. As a reference, the
mean mass in the range $13,000 < T_{\rm eff}$ (K) $< 40,000$ is shown by red
circles and dotted lines. We see that in terms of the mean mass, both samples
show a similar variation as a function of $T_{\rm eff}$. In the 3D case, there
is a clear increase in the $11,000 < T_{\rm eff}$ (K) $< 12,000$ bin for both
surveys. We currently have no explanation for this behaviour. We have verified
that changes in the parameterisation of the line core removal do little to
improve the situation. While the reason might be the inaccuracy of our first
grid of 3D model atmospheres, we could also be observing some issues with the
spectral synthesis, e.g. from the opacities or EOS, that were previously
hidden behind the high-$\log g$ problem.

The variation of the mass standard deviation as a function of $T_{\rm eff}$
(bottom panel of Fig.~\ref{fg:f_bin}) has a rather complex behaviour. The
overall lower values in the 3D case simply reflect the fact that the
gravities are smaller, hence also the mass range. The increase of the
dispersion at low $T_{\rm eff}$, especially in the case of the SDSS sample, is
likely related to the fact that lines are weaker and that the mean $\log g$
error increases. The minima in the standard deviation at $T_{\rm eff} \sim$
11,500-12,500~K are more difficult to explain. This could be due to incorrect
3D $T_{\rm eff}$ corrections, especially for the low- and high-mass objects
which could be systematically shifted to another bin.

For cool convective white dwarfs ($T_{\rm eff} \lesssim 8000$~K), neutral
broadening and related non-ideal effects due to neutral particles become
dominant for the Balmer lines. In their discussion of the high-$\log g$
problem, \citet{TB10} mention how the uncertainties in neutral broadening and
the \citet{hm88} theory for non-ideal perturbations (hard sphere model) may
impact the model predictions. In particular, \citet{bergeron91} found that
$\log g$ values are too low in the regime where non-ideal effects become
dominated by neutral interactions, hence they divided the hydrogen radius by a
factor of 2 in the Hummer \& Mihalas hard sphere formalism. While it is now
clear that these uncertainties are not the source of the high-$\log g$
problem, they may impact the spectroscopic mass distributions for $T_{\rm eff}
\lesssim 8000$~K. The 3D mass distributions in Fig.~\ref{fg:f_bin} remain
relatively stable as a function of $T_{\rm eff}$ for the coolest white dwarfs.
Our results suggest that the current parameterisation of the non-ideal effects
due to neutral interactions does not need to be revised for the 3D
effects. Finally, it is worth mentionning that our assumption of thick H
layers for all DA white dwarfs also has a slight effect on our
results. Considering that most ($\gtrsim 80\%$) DAs appear to have thick
layers \citep{TB08}, we expect an offset below $1\%$ on the derived mean
masses.

\subsection{ZZ Ceti white dwarfs}

We present in this section a brief study of the pulsating ZZ Ceti properties
as seen in the light of 3D model atmospheres. Our starting point is once again
the sample of GB11 for which they made a distinction between pulsating,
photometrically constant and unconstrained objects. In the latter case, the
stars were not observed to detect possible light variations. In their Fig.~33,
GB11 have studied the position of the ZZ Ceti instability strip relying only
on the stars in their sample that have been observed for photometric
variability. Their purpose was to update the position of the instability strip
given the improved 1D model atmospheres of \citet{TB09} and the resulting
change in the MLT calibration from ML2/$\alpha$ = 0.6 to 0.8.  We propose here
a very similar study with our improved 3D model atmospheres. 


\begin{figure}[!h]
\begin{center}
\includegraphics[bb=38 172 572 642,width=3.2in]{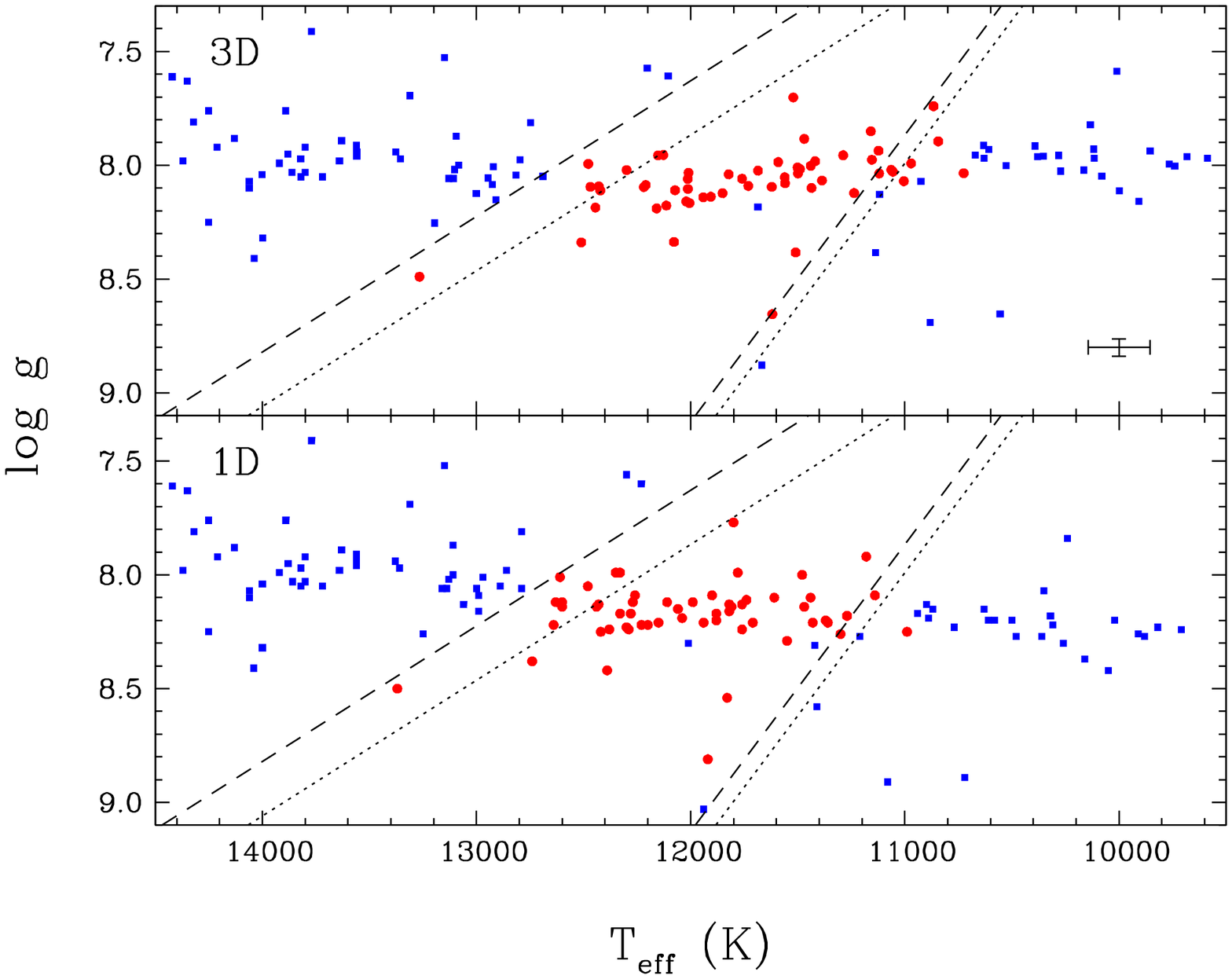}
\caption{Surface gravity versus $T_{\rm eff}$ distribution for ZZ Ceti
  (circles, red) and photometrically constant (squares, blue) DA white dwarfs
  in the \citet{gianninas11} sample with $\langle$3D$\rangle$ (top) and 1D
  (bottom) spectra. The dotted lines are the ZZ Ceti instability strip
  boundaries established by \citet{gianninas06} while the dashed lines are the
  most recent boundaries identified by \citet{gianninas11} relying on the
  models from \citet{TB09}. The error bars represent the average uncertainties
  of the spectroscopic method in the region of the ZZ Ceti instability strip.
\label{fg:f_ZZ}}
\end{center}
\end{figure}

In Fig.\ref{fg:f_ZZ}, we present the $\log g$ distribution of the GB11 sample
in the region of the ZZ Ceti instability strip and identify photometrically
variable and constant objects. For reference, we also add the {\it observed}
red and blue edges of the instability strip as determined by GB11
(ML2/$\alpha$ = 0.8, dashed lines). We remind the reader that while the
  observed edges depend on the predicted Balmer lines and the model
  atmospheres, the MLT parameterisation (see Sect.~3.3) is independent of
  asteroseismic predictions. The $T_{\rm eff}$ and $\log g$ shifts are
moderate in this regime, hence the position and shape of the instability strip
with $\langle$3D$\rangle$ spectra are not dramatically different in comparison
to the 1D ML2/$\alpha$ = 0.8 parameterisation. The blue and red edge appear to
be slightly cooler. Furthermore, the purity of the instability strip and the
sharpness of the edges are not modified because the 3D corrections have a
largely systematic behaviour.

We note that recent {\it theoretical} determinations of the blue and red edge
positions from asteroseismology
\citep{fontaine08,corsico12,grootel12,grootel13} are in agreement with the
GB11 or earlier \citet{gianninas06} {\it observed} positions. It is
  encouraging that the observed 3D boundaries are also in agreement with
  asteroseismology, although 3D $T_{\rm eff}$ corrections around $12,500$~K
are sensitive to the line cores, hence our results for ZZ Ceti stars should be
taken with some caution. Furthermore, asteroseismic models rely on 1D
structure models (typically under the ML2/$\alpha$ = 1.0 parameterisation) and
it is currently unknown how much the 3D model atmospheres would impact the
results. In addition to different convective efficiencies between the 1D
  and 3D models leading to distinctive sizes for the convective zone, the 3D
  simulations show overshoot layers at the bottom of the convective zone which
  are missing from the 1D models.

\section{Conclusion}

We presented a new grid of spectra for DA white dwarfs drawn from CO$^5$BOLD
3D radiation-hydrodynamics simulations. The 3D effects on the atmospheric
parameters of DA stars have been studied for the SDSS and White Dwarf Catalog
spectroscopic samples \citep{TB11,gianninas11}. We found that the mean
spectroscopic mass of cool convective white dwarfs is significantly lower, by
about $\sim$0.10 $M{_\odot}$, when we rely on 3D instead of 1D model
atmospheres. The mean mass of objects with hot radiative and cool convective
atmospheres is now roughly the same for both spectroscopic samples. This is in
much better agreement with independent parallax observations and our knowledge
of stellar evolution. We have proposed correction functions that can be used
to convert spectroscopically determined 1D atmospheric parameters to 3D
values.  The 3D atmospheric parameters can then be utilised to determine
improved values of the mass, age, and population membership for convective DA
white dwarfs.

We found that 3D $T_{\rm eff}$ corrections are small, and that they are in
agreement with photometric and near-UV determined $T_{\rm eff}$ values. We
nevertheless noted some shortcomings of the current grid of
$\langle$3D$\rangle$ spectra. The masses are predicted slightly too high
($\sim$5\%) at $T_{\rm eff} \sim 12,000$~K and $T_{\rm eff}$ corrections are
also uncertain in this regime ($\pm$250~K). These issues might not only be
related to the 3D simulations, but also to shortcomings in the predicted line
profiles that were previously hidden behind the high-$\log g$ problem.

We demonstrated that the observed ZZ Ceti instability strip remains at a very
similar position in the HR diagram compared to previous determinations with 1D
spectra. We hope, however, that our 3D atmospheres can also be implemented as
the upper boundary of white dwarf structure calculations, for instance by
using the methods of \citet{gautschy96} and \citet{ludwig99}. This would allow
the asteroseismic applications to also benefit from an improved model of
convection, and to determine whether there is an impact on the theoretical
position of the ZZ Ceti instability strip.

\begin{acknowledgements}
P.-E. T. is supported by the Alexander von Humboldt Foundation. 3D model
calculations have been performed on CALYS, a mini-cluster of 320 nodes built
at Universit\'e de Montr\'eal with the financial help of the Fondation
Canadienne pour l'Innovation. We thank Prof. G. Fontaine for making CALYS
available to us. We are also most grateful to Dr. P. Brassard for technical
help. This work was supported by Sonderforschungsbereich SFB 881 "The Milky
Way System" (Subproject A4) of the German Research Foundation
(DFG). B.F.\ acknowledges financial support from the {\sl Agence Nationale de
  la Recherche} (ANR), and the {\sl ``Programme Nationale de Physique
  Stellaire''} (PNPS) of CNRS/INSU, France.
\end{acknowledgements}

\bibliographystyle{aa} 

\Online

\begin{appendix} 

 \begin{table*}[h!]
\section{Supplementary data I}
 \caption{Grid of CO$^5$BOLD 3D model atmospheres for DA white dwarfs}
 \label{tab1}
 \begin{center}
 \begin{tabular}{llllllllllll}
\hline
\hline
$T_{\rm eff}$ & $\log g$ & $\log z$ & $\log x$ & $\log$ Char. size & $\log
\tau_{\rm R, min}$ & $\log
\tau_{\rm R, max}$ & $\ln(P_{\rm bot}/P_{\rm
  phot})$ & $t$ & $\log t_{\rm adv}$ & $\log t_{\rm
  decay}$ & $\delta I_{\rm rms}/\langle I \rangle$ \\

(K) & & [cm] & [cm] & [cm] & & & & (s) & [s] & [s] & (\%) \\
\hline
     6112&  7.00&  5.58&  6.01&  4.61& -8.61&  3.00&  5.50& 100.0& -0.41& -0.08&  3.55\\
     7046&  7.00&  5.66&  6.06&  4.73& -9.33&  3.00&  5.15& 100.0& -0.61& -0.02&  9.27\\
     8027&  7.00&  5.78&  6.10&  4.83& -7.63&  3.01&  5.11& 100.0& -0.70& -0.06& 15.62\\
     9025&  7.00&  5.89&  6.23&  5.01& -6.09&  2.99&  3.30& 100.0& -0.79& -0.19& 19.07\\
     9521&  7.00&  6.01&  6.40&  5.07& -5.65&  3.01&  3.17& 100.0& -0.83& -0.11& 19.37\\
    10018&  7.00&  6.11&  6.40&  5.18& -4.96&  2.75&  2.50& 100.0& -0.92& -0.06& 19.31\\
    10540&  7.00&  6.75&  6.87&  5.30& -6.08&  3.11&  1.66&  50.0& -0.91& -0.28& 18.90\\
    11000&  7.00&  6.75&  6.87&  5.40& -5.07&  3.00&  1.33&  50.0& -0.83& -0.21& 19.39\\
    11501&  7.00&  6.81&  6.87&  5.30& -6.56&  3.00&  1.79&  50.0& -0.63& -0.29& 13.03\\
    12001&  7.00&  6.85&  6.87&  5.40& -6.59&  3.00&  1.91&  50.0& -0.41& -0.43&  6.90\\
    12501&  7.00&  6.88&  6.87&  5.40& -6.12&  3.00&  1.91&  50.0& -0.17& -0.49&  2.94\\
    13003&  7.00&  6.90&  6.87&  5.39& -6.41&  3.00&  2.11&  50.0&  0.05& -0.46&  1.39\\
     6065&  7.50&  5.07&  5.56&  4.08& -8.33&  2.99&  5.12&  31.6& -0.78& -0.44&  1.83\\
     7033&  7.50&  5.17&  5.60&  4.20& -9.73&  3.00&  5.28&  31.6& -0.96& -0.49&  5.90\\
     8017&  7.50&  5.27&  5.60&  4.33& -8.66&  3.00&  5.24&  31.6& -1.11& -0.46& 11.21\\
     9015&  7.50&  5.37&  5.68&  4.46& -7.21&  3.00&  4.02&  31.6& -1.19& -0.47& 15.83\\
     9549&  7.50&  5.49&  5.78&  4.51& -6.25&  3.01&  3.81&  31.6& -1.25& -0.57& 17.02\\
    10007&  7.50&  5.56&  5.78&  4.65& -6.01&  3.00&  3.58&  31.6& -1.26& -0.63& 17.32\\
    10500&  7.50&  5.66&  5.84&  4.70& -5.43&  2.99&  2.77&  31.6& -1.31& -0.58& 17.86\\
    10938&  7.50&  5.98&  6.25&  4.77& -5.44&  2.90&  2.19&  31.6& -1.40& -0.60& 19.16\\
    11498&  7.50&  6.30&  6.33&  5.00& -5.77&  3.01&  1.65&  31.6& -1.34& -0.50& 19.85\\
    11999&  7.50&  6.33&  6.33&  4.93& -6.12&  3.01&  1.68&  31.6& -1.20& -0.67& 17.68\\
    12500&  7.50&  6.37&  6.33&  4.85& -6.19&  3.00&  1.89&  31.6& -1.00& -0.86& 10.73\\
    13002&  7.50&  6.41&  6.33&  4.85& -6.81&  3.00&  2.21&  31.6& -0.85& -1.05&  6.08\\
     5997&  8.00&  4.56&  5.08&  3.60& -8.51&  3.00&  5.31&  10.0& -1.09& -1.06&  0.87\\
     7011&  8.00&  4.65&  5.08&  3.75& -8.70&  3.00&  4.50&  10.0& -1.36& -1.07&  3.54\\
     8034&  8.00&  4.74&  5.15&  3.88& -8.77&  3.00&  4.01&  10.0& -1.50& -1.01&  7.58\\
     9036&  8.00&  4.84&  5.20&  3.93& -7.72&  2.99&  3.49&  10.0& -1.58& -0.97& 11.98\\
     9518&  8.00&  4.96&  5.32&  3.99& -6.27&  3.01&  3.56&  10.0& -1.64& -1.08& 13.57\\
    10025&  8.00&  5.03&  5.32&  4.05& -6.74&  3.01&  4.17&  10.0& -1.70& -1.11& 14.43\\
    10532&  8.00&  5.10&  5.35&  4.13& -6.37&  3.01&  3.61&  10.0& -1.74& -1.09& 15.00\\
    11005&  8.00&  5.20&  5.43&  4.21& -6.21&  3.00&  4.13&  10.0& -1.78& -1.10& 16.61\\
    11529&  8.00&  5.37&  5.57&  4.29& -6.05&  3.00&  3.61&  10.0& -1.80& -1.06& 17.74\\
    11980&  8.00&  5.93&  5.88&  4.40& -6.62&  3.28&  2.25&  10.0& -1.81& -0.99& 18.72\\
    12099&  8.00&  5.89&  5.88&  4.48& -6.72&  3.12&  2.23&  10.0& -1.79& -0.96& 18.96\\
    12504&  8.00&  5.87&  5.88&  4.54& -6.25&  3.00&  1.92&  10.0& -1.72& -0.95& 19.21\\
    13000&  8.00&  5.90&  5.88&  4.48& -6.23&  3.01&  1.95&  10.0& -1.57& -1.15& 16.00\\
    13502&  8.00&  5.73&  5.88&  4.40& -5.51&  2.49&  2.38&  10.0& -1.41& -1.30&  9.32\\
    14000&  8.00&  5.75&  5.88&  4.40& -4.97&  2.47&  2.42&  12.5& -1.25& -1.37&  4.63\\
     6024&  8.50&  4.09&  4.57&  3.09& -8.66&  3.00&  5.80&   3.2& -1.50& -1.40&  0.54\\
     6925&  8.50&  4.14&  4.65&  3.17& -8.66&  3.00&  4.57&   3.2& -1.67& -1.82&  1.90\\
     8004&  8.50&  4.24&  4.65&  3.25& -8.81&  3.00&  3.99&   3.2& -1.89& -1.42&  4.74\\
     9068&  8.50&  4.34&  4.69&  3.36& -9.21&  3.00&  3.73&   3.2& -2.00& -1.43&  8.42\\
     9522&  8.50&  4.43&  4.74&  3.47& -7.43&  2.99&  4.17&   3.2& -2.05& -1.50& 10.06\\
     9972&  8.50&  4.50&  4.77&  3.50& -6.42&  3.00&  3.69&   3.2& -2.09& -1.50& 11.22\\
    10496&  8.50&  4.54&  4.80&  3.53& -6.38&  3.00&  3.28&   3.2& -2.14& -1.46& 12.10\\
    10997&  8.50&  4.65&  4.92&  3.65& -6.23&  3.00&  3.32&   3.2& -2.18& -1.48& 13.37\\
    11490&  8.50&  4.75&  5.07&  3.74& -6.67&  3.00&  3.96&   3.2& -2.20& -1.61& 14.26\\
    11979&  8.50&  4.83&  5.10&  3.82& -6.35&  3.00&  3.67&   3.2& -2.22& -1.51& 15.18\\
    12420&  8.50&  4.96&  5.12&  3.85& -6.09&  3.00&  3.47&   3.2& -2.21& -1.56& 16.31\\
    12909&  8.50&  5.35&  5.29&  3.96& -5.54&  2.96&  1.79&   3.2& -2.19& -1.48& 17.07\\
    13453&  8.50&  5.38&  5.38&  4.04& -5.06&  2.94&  1.79&   4.0& -2.15& -1.44& 18.48\\
    14002&  8.50&  5.42&  5.38&  3.98& -5.47&  2.95&  1.97&   4.0& -2.02& -1.54& 14.62\\
    14492&  8.50&  5.43&  5.38&  3.90& -5.38&  2.92&  2.22&   4.0& -1.82& -1.68&  9.23\\
\hline
\end{tabular} 
\end{center} 
\tablefoot{All quantities were extracted and temporally averaged from at least
  250 snapshots in the last half of the simulation. $T_{\rm eff}$ is derived
  from the temporal and spatial average of the emergent flux. The
  characteristic size, decay time and relative intensity contrast ($\delta I_{\rm
    rms}/\langle I \rangle$) of the granulation were computed from intensity maps using the
  definitions given in \citet{tremblay13b}. $P_{\rm bot}$ is the pressure at
  the bottom layer, $P_{\rm phot}$ the pressure at $\tau_{\rm R} = 1$, and the
  computation time $t$ is in stellar time not including the initial 2D runs
  for cooler models. The turnover or advective timescale ($t_{\rm adv}$) is evaluated at
  $\tau_{\rm R} = 1$. For depth dependent quantities, the spatial average
  was performed over constant Rosseland optical depth.}
\end{table*}

\setcounter{table}{0}
 \begin{table*}
 \label{tab1}
 \begin{center}
 \caption{continued}
 \begin{tabular}{llllllllllll}
\hline
\hline
$T_{\rm eff}$ & $\log g$ & $\log z$ & $\log x$ & $\log$ Char. size & $\log
\tau_{\rm R, min}$ & $\log
\tau_{\rm R, max}$ & $\ln(P_{\rm bot}/P_{\rm
  phot})$ & $t$ & $\log t_{\rm adv}$ & $\log t_{\rm
  decay}$ & $\delta I_{\rm rms}$ \\

(K) & & [cm] & [cm] & [cm] & & & & (s) & [s] & [s] & (\%) \\
\hline
     6028&  9.00&  3.57&  4.12&  2.64& -8.32&  2.99&  4.75&   1.0& -1.85& -2.18&  0.33\\
     6960&  9.00&  3.67&  4.12&  2.64& -7.85&  2.96&  4.02&   1.0& -2.08& -2.12&  1.10\\
     8041&  9.00&  3.80&  4.16&  2.76& -9.18&  2.99&  4.29&   1.0& -2.31& -1.91&  2.97\\
     8999&  9.00&  3.70&  4.16&  2.89& -8.70&  2.98&  2.05&   1.0& -2.39& -1.93&  5.39\\
     9507&  9.00&  3.87&  4.24&  2.91& -6.02&  3.00&  3.03&   1.0& -2.43& -1.91&  6.79\\
     9962&  9.00&  3.97&  4.27&  2.94& -8.08&  3.00&  4.15&   1.0& -2.46& -1.95&  7.97\\
    10403&  9.00&  4.01&  4.30&  3.03& -7.51&  3.00&  4.19&   1.0& -2.51& -1.93&  8.92\\
    10948&  9.00&  4.10&  4.37&  3.04& -6.27&  3.00&  3.72&   1.0& -2.53& -1.98& 10.09\\
    11415&  9.00&  4.11&  4.37&  3.09& -5.91&  2.99&  3.33&   1.0& -2.57& -1.97& 10.77\\
    11915&  9.00&  4.25&  4.55&  3.22& -7.02&  3.01&  3.85&   1.0& -2.62& -1.98& 11.70\\
    12436&  9.00&  4.32&  4.57&  3.30& -6.96&  3.00&  3.94&   1.0& -2.64& -2.00& 12.53\\
    12969&  9.00&  4.44&  4.69&  3.36& -6.63&  3.00&  4.09&   1.0& -2.68& -1.92& 13.65\\
    13496&  9.00&  4.61&  4.87&  3.48& -6.50&  2.79&  3.48&   1.0& -2.69& -1.95& 15.20\\
    14008&  9.00&  4.96&  4.87&  3.48& -5.86&  3.10&  1.98&   1.0& -2.65& -1.81& 16.69\\
    14591&  9.00&  4.90&  4.97&  3.57& -4.80&  2.89&  1.84&   1.2& -2.53& -1.85& 16.67\\
    14967&  9.00&  4.96&  4.88&  3.48& -5.69&  2.91&  2.30&   1.2& -2.44& -1.92& 14.43\\
\hline
\end{tabular} 
\end{center} 
\end{table*}

\end{appendix}

\begin{appendix} 

 \begin{table*}[h!]
\section{Supplementary data II}
 \caption{1D ML2/$\alpha$ = 0.8 to 3D $T_{\rm eff}$ corrections}
 \label{tab3}
 \begin{center}
 \begin{tabular}{llllll}
\hline
\hline
$T_{\rm eff}$ (K) & $\log g$ = 7.0 & $\log g$ = 7.5 & $\log g$ = 8.0 & $\log g$ = 8.5 & $\log g$ = 9.0 \\
\hline
  6000 &     13 &     15 &      9 &     $-$7 &    $-$23 \\
  6500 &     13 &     27 &     27 &     14 &      4 \\
  7000 &      3 &     19 &     12 &    $-$13 &     18 \\
  7500 &    $-$25 &      0 &      5 &     $-$3 &    $-$50 \\
  8000 &    $-$67 &    $-$21 &      2 &     $-$4 &    $-$39 \\
  8500 &   $-$103 &    $-$63 &    $-$22 &     $-$1 &    $-$51 \\
  9000 &   $-$154 &   $-$113 &    $-$68 &    $-$29 &    $-$51 \\
  9500 &   $-$219 &   $-$162 &   $-$136 &    $-$98 &    $-$60 \\
 10000 &   $-$293 &   $-$205 &   $-$180 &   $-$131 &    $-$77 \\
 10500 &   $-$474 &   $-$292 &   $-$235 &   $-$178 &   $-$124 \\
 11000 &   $-$416 &   $-$348 &   $-$236 &   $-$216 &   $-$133 \\
 11500 &   $-$277 &   $-$362 &   $-$260 &   $-$240 &   $-$241 \\
 12000 &    $-$34 &   $-$150 &   $-$321 &   $-$301 &   $-$472 \\
 12500 &      0 &     36 &   $-$158 &   $-$296 &   $-$456 \\
 13000 &      0 &      0 &     39 &   $-$305 &   $-$321 \\
 13500 &      0 &      0 &    166 &    $-$47 &   $-$348 \\
 14000 &      0 &      0 &      0 &      4 &   $-$187 \\
 14500 &      0 &      0 &      0 &      0 &    $-$44 \\
\hline
\end{tabular} 
\end{center} 
\end{table*}

 \begin{table*}[h!]
 \caption{1D ML2/$\alpha$ = 0.8 to 3D $\log g$ corrections}
 \label{tab3}
 \begin{center}
 \begin{tabular}{llllll}
\hline
\hline
$T_{\rm eff}$ (K) & $\log g$ = 7.0 & $\log g$ = 7.5 & $\log g$ = 8.0 & $\log g$ = 8.5 & $\log g$ = 9.0 \\
\hline
  6000 & $-$.081 & $-$.019 & 0.014 & 0.037 & 0.053 \\
  6500 & $-$.076 & $-$.028 & $-$.030 & $-$.001 & 0.028 \\
  7000 & $-$.154 & $-$.100 & $-$.074 & $-$.084 & $-$.023 \\
  7500 & $-$.195 & $-$.159 & $-$.135 & $-$.108 & $-$.150 \\
  8000 & $-$.224 & $-$.189 & $-$.166 & $-$.164 & $-$.160 \\
  8500 & $-$.256 & $-$.255 & $-$.244 & $-$.223 & $-$.225 \\
  9000 & $-$.303 & $-$.281 & $-$.277 & $-$.271 & $-$.277 \\
  9500 & $-$.288 & $-$.261 & $-$.260 & $-$.268 & $-$.316 \\
 10000 & $-$.256 & $-$.240 & $-$.244 & $-$.272 & $-$.309 \\
 10500 & $-$.184 & $-$.202 & $-$.221 & $-$.269 & $-$.210 \\
 11000 & 0.035 & $-$.121 & $-$.146 & $-$.208 & $-$.180 \\
 11500 & 0.051 & $-$.040 & $-$.106 & $-$.147 & $-$.192 \\
 12000 & 0.031 & 0.015 & $-$.039 & $-$.105 & $-$.114 \\
 12500 & 0.0 & 0.048 & $-$.030 & $-$.055 & $-$.093 \\
 13000 & 0.0 & 0.0 & $-$.008 & $-$.039 & $-$.067 \\
 13500 & 0.0 & 0.0 & 0.021 & $-$.006 & $-$.004 \\
 14000 & 0.0 & 0.0 & 0.0 & 0.018 & 0.011 \\
 14500 & 0.0 & 0.0 & 0.0 & 0.0 & 0.025 \\
\hline
\end{tabular} 
\end{center} 
\end{table*}

 \begin{table*}[h!]
 \caption{1D ML2/$\alpha$ = 0.7 to 3D $T_{\rm eff}$ corrections}
 \label{tab3}
 \begin{center}
 \begin{tabular}{llllll}
\hline
\hline
$T_{\rm eff}$ (K) & $\log g$ = 7.0 & $\log g$ = 7.5 & $\log g$ = 8.0 & $\log g$ = 8.5 & $\log g$ = 9.0 \\
\hline
  6000 &     15 &     16 &      9 &     $-$7 &    $-$23 \\
  6500 &     20 &     31 &     28 &     23 &      5 \\
  7000 &     13 &     25 &     16 &    $-$12 &     19 \\
  7500 &     $-$8 &      9 &     11 &      0 &    $-$48 \\
  8000 &    $-$39 &     $-$4 &     11 &      1 &    $-$37 \\
  8500 &    $-$64 &    $-$38 &     $-$9 &      6 &    $-$47 \\
  9000 &   $-$101 &    $-$75 &    $-$45 &    $-$17 &    $-$45 \\
  9500 &   $-$143 &   $-$104 &    $-$98 &    $-$77 &    $-$51 \\
 10000 &   $-$172 &   $-$119 &   $-$116 &    $-$95 &    $-$60 \\
 10500 &   $-$295 &   $-$148 &   $-$134 &   $-$111 &    $-$89 \\
 11000 &   $-$232 &   $-$154 &    $-$66 &    $-$99 &    $-$68 \\
 11500 &   $-$115 &   $-$146 &    $-$39 &    $-$30 &    $-$91 \\
 12000 &     69 &     76 &    $-$75 &    $-$36 &   $-$194 \\
 12500 &      0 &    226 &     73 &    $-$22 &   $-$136 \\
 13000 &      0 &      0 &    289 &    $-$37 &    $-$22 \\
 13500 &      0 &      0 &    402 &    214 &    $-$79 \\
 14000 &      0 &      0 &      0 &    288 &     66 \\
 14500 &      0 &      0 &      0 &      0 &    265 \\
\hline
\end{tabular} 
\end{center} 
\end{table*}

 \begin{table*}[h!]
 \caption{1D ML2/$\alpha$ = 0.7 to 3D $\log g$ corrections}
 \label{tab3}
 \begin{center}
 \begin{tabular}{llllll}
\hline
\hline
$T_{\rm eff}$ (K) & $\log g$ = 7.0 & $\log g$ = 7.5 & $\log g$ = 8.0 & $\log g$ = 8.5 & $\log g$ = 9.0 \\
\hline
  6000 & $-$.078 & $-$.020 & 0.014 & 0.037 & 0.053 \\
  6500 & $-$.072 & $-$.026 & $-$.029 & 0.016 & 0.028 \\
  7000 & $-$.149 & $-$.097 & $-$.072 & $-$.083 & $-$.023 \\
  7500 & $-$.190 & $-$.156 & $-$.132 & $-$.107 & $-$.150 \\
  8000 & $-$.217 & $-$.186 & $-$.164 & $-$.162 & $-$.159 \\
  8500 & $-$.248 & $-$.253 & $-$.242 & $-$.222 & $-$.224 \\
  9000 & $-$.300 & $-$.281 & $-$.276 & $-$.270 & $-$.275 \\
  9500 & $-$.289 & $-$.265 & $-$.263 & $-$.268 & $-$.316 \\
 10000 & $-$.266 & $-$.251 & $-$.253 & $-$.277 & $-$.309 \\
 10500 & $-$.194 & $-$.223 & $-$.240 & $-$.279 & $-$.215 \\
 11000 & 0.035 & $-$.140 & $-$.174 & $-$.230 & $-$.192 \\
 11500 & 0.064 & $-$.064 & $-$.135 & $-$.181 & $-$.218 \\
 12000 & 0.044 & 0.0 & $-$.070 & $-$.145 & $-$.160 \\
 12500 & 0.0   & 0.042 & $-$.061 & $-$.092 & $-$.142 \\
 13000 & 0.0   & 0.0 & $-$.038 & $-$.077 & $-$.113 \\
 13500 & 0.0   & 0.0 & $-$.002 & $-$.042 & $-$.043 \\
 14000 & 0.0   & 0.0 & 0.0 & $-$.013 & $-$.025 \\
 14500 & 0.0   & 0.0 & 0.0 & 0.0 & $-$.010 \\
\hline
\end{tabular} 
\end{center} 
\end{table*}

 \begin{table*}[h!]
 \caption{1D ML2/$\alpha$ = 0.6 to 3D $T_{\rm eff}$ corrections}
 \label{tab3}
 \begin{center}
 \begin{tabular}{llllll}
\hline
\hline
$T_{\rm eff}$ (K) & $\log g$ = 7.0 & $\log g$ = 7.5 & $\log g$ = 8.0 & $\log g$ = 8.5 & $\log g$ = 9.0 \\
\hline
  6000 &     17 &     16 &      9 &     $-$7 &    $-$23 \\
  6500 &     28 &     35 &     31 &     24 &      6 \\
  7000 &     27 &     33 &     22 &    $-$10 &     19 \\
  7500 &     15 &     22 &     18 &      4 &    $-$46 \\
  8000 &     $-$4 &     17 &     22 &      7 &    $-$33 \\
  8500 &    $-$15 &     $-$7 &      9 &     16 &    $-$41 \\
  9000 &    $-$40 &    $-$30 &    $-$18 &     $-$2 &    $-$37 \\
  9500 &    $-$59 &    $-$38 &    $-$51 &    $-$51 &    $-$38 \\
 10000 &    $-$51 &    $-$29 &    $-$46 &    $-$52 &    $-$39 \\
 10500 &   $-$112 &     $-$9 &    $-$31 &    $-$40 &    $-$48 \\
 11000 &    $-$55 &     45 &     89 &     14 &     $-$1 \\
 11500 &     78 &     85 &    181 &    148 &     37 \\
 12000 &    162 &    316 &    192 &    223 &     72 \\
 12500 &      0 &    449 &    360 &    285 &    203 \\
 13000 &      0 &      0 &    578 &    291 &    325 \\
 13500 &      0 &      0 &    696 &    500 &    238 \\
 14000 &      0 &      0 &      0 &    604 &    351 \\
 14500 &      0 &      0 &      0 &      0 &    535 \\
\hline
\end{tabular} 
\end{center} 
\end{table*}

 \begin{table*}[h!]
 \caption{1D ML2/$\alpha$ = 0.6 to 3D $\log g$ corrections}
 \label{tab3}
 \begin{center}
 \begin{tabular}{llllll}
\hline
\hline
$T_{\rm eff}$ (K) & $\log g$ = 7.0 & $\log g$ = 7.5 & $\log g$ = 8.0 & $\log g$ = 8.5 & $\log g$ = 9.0 \\
\hline
  6000 & $-$.075 & $-$.021 & 0.014 & 0.037 & 0.053 \\
  6500 & $-$.067 & $-$.023 & $-$.028 & 0.016 & 0.029 \\
  7000 & $-$.143 & $-$.094 & $-$.067 & $-$.082 & $-$.023 \\
  7500 & $-$.181 & $-$.152 & $-$.129 & $-$.105 & $-$.149 \\
  8000 & $-$.204 & $-$.181 & $-$.160 & $-$.160 & $-$.157 \\
  8500 & $-$.234 & $-$.249 & $-$.239 & $-$.219 & $-$.222 \\
  9000 & $-$.293 & $-$.278 & $-$.276 & $-$.268 & $-$.274 \\
  9500 & $-$.285 & $-$.267 & $-$.264 & $-$.269 & $-$.314 \\
 10000 & $-$.269 & $-$.261 & $-$.260 & $-$.280 & $-$.310 \\
 10500 & $-$.199 & $-$.240 & $-$.257 & $-$.289 & $-$.219 \\
 11000 & 0.042 & $-$.157 & $-$.199 & $-$.250 & $-$.204 \\
 11500 & 0.079 & $-$.081 & $-$.165 & $-$.212 & $-$.239 \\
 12000 & 0.060 & $-$.020 & $-$.101 & $-$.183 & $-$.200 \\
 12500 & 0.0 & 0.036 & $-$.095 & $-$.135 & $-$.193 \\
 13000 & 0.0 & 0.0 & $-$.072 & $-$.120 & $-$.165 \\
 13500 & 0.0 & 0.0 & $-$.029 & $-$.082 & $-$.090 \\
 14000 & 0.0 & 0.0 & 0.0 & $-$.051 & $-$.068 \\
 14500 & 0.0 & 0.0 & 0.0 & 0.0 & $-$.051 \\
\hline
\end{tabular} 
\end{center} 
\end{table*}

\end{appendix}

\onecolumn

\begin{appendix} 

\section{Correction functions for Fortran 77}

\lstset{
    breaklines     = true,
    numbers        = left,
    stepnumber     = 1,
}
\lstset{language=Fortran}

\subsection{1D ML2/$\alpha$ = 0.8 to 3D $T_{\rm eff}$ corrections}

\begin{lstlisting}
      function ML18_to_3D_dTeff(Teff,logg)
      real*8 Teff, logg, a(8), Shift, Teff0, logg0, ML18_to_3D_dTeff
c 
c     IN: Teff, effective temperature (K)
c         logg, surface gravity (g in cm/s2)
c
c     OUT: Teff correction (K)
c      ; Fri Jun 14 20:16:39 2013 B.F.
      
      A(1)=1.0947335E-03
      A(2)=-1.8716231E-01
      A(3)=1.9350009E-02
      A(4)=6.4821613E-01
      A(5)=-2.2863187E-01
      A(6)=5.8699232E-01
      A(7)=-1.0729871E-01
      A(8)=1.1009070E-01

      Teff0=(Teff-10000.0)/1000.00
      logg0=(logg-8.00000)/1.00000

      Shift=A(1)+(A(2)+A(7)*Teff0+A(8)*logg0)*exp(-(A(3)+A(5)*Teff0+
     *A(6)*logg0)**2*((Teff0-A(4))**2))

      ML18_to_3D_dTeff = Shift*1000.00+0.00000
      return
      end 
\end{lstlisting}
~\\

\subsection{1D ML2/$\alpha$ = 0.8 to 3D $\log g$ corrections}

\begin{lstlisting}
      function ML18_to_3D_dlogg(Teff,logg)
      real*8 Teff, logg, a(12), Shift,Teff0,logg0,ML18_to_3D_dlogg
c 
c     IN: Teff, effective temperature (K)
c         logg, surface gravity (g in cm/s2)
c
c     OUT: log g correction (g in cm/s2)
c    ; Fri Jun 14 21:09:56 2013 B.F.

      A(1)=7.5209868E-04
      A(2)=-9.2086619E-01
      A(3)=3.1253746E-01
      A(4)=-1.0348176E+01
      A(5)=6.5854716E-01
      A(6)=4.2849862E-01
      A(7)=-8.8982873E-02
      A(8)=1.0199718E+01
      A(9)=4.9277883E-02
      A(10)=-8.6543477E-01
      A(11)=3.6232756E-03
      A(12)=-5.8729354E-02

      Teff0=(Teff-10000.0)/1000.00
      logg0=(logg-8.00000)/1.00000
      Shift=(A(1)+A(5)*exp(-A(6)*((Teff0-A(7))**2)))+A(2)*
     *exp(-A(3)*((Teff0-(A(4)+A(8)*exp(-(A(9)+A(11)*
     *Teff0+A(12)*logg0)**2*((Teff0-A(10))**2))))**2))

      ML18_to_3D_dlogg = Shift
      return
      end 
\end{lstlisting}
~\\

\subsection{1D ML2/$\alpha$ = 0.7 to 3D $T_{\rm eff}$ corrections}

\begin{lstlisting}
      function ML17_to_3D_dTeff(Teff,logg)
      real*8 Teff, logg, a(8), Shift, Teff0, logg0, ML17_to_3D_dTeff
c 
c     IN: Teff, effective temperature (K)
c         logg, surface gravity (g in cm/s2)
c
c     OUT: Teff correction (K)
c    ; Tue Jun 18 18:58:49 2013 B.F.

      A(1)=-1.0461690E-03
      A(2)=-2.6846737E-01
      A(3)=3.0654611E-01
      A(4)=1.8025848E+00
      A(5)=1.5006909E-01
      A(6)=1.0125295E-01
      A(7)=-5.2933335E-02
      A(8)=-1.3414353E-01

      Teff0=(Teff-10000.0)/1000.00
      logg0=(logg-8.00000)/1.00000
      Shift=A(1)+(A(2)+A(5)*Teff0+(A(6)+A(7)*Teff0+A(8)*logg0)*logg0)*
     *exp(-A(3)*((Teff0-A(4))**2))

      ML17_to_3D_dTeff = Shift*1000.00+0.00000
      return
      end 

\end{lstlisting}
~\\
\subsection{1D ML2/$\alpha$ = 0.7 to 3D $\log g$ corrections}

\begin{lstlisting}
      function ML17_to_3D_dlogg(Teff,logg)
      real*8 Teff, logg, a(11), Shift, Teff0, logg0, ML17_to_3D_dlogg
c 
c     IN: Teff, effective temperature (K)
c         logg, surface gravity (g in cm/s2)
c
c     OUT: log g correction (g in cm/s2)
c    ; Tue Jun 18 18:56:55 2013 B.F.

      A(1)= 1.1922481E-03
      A(2)=-2.7230889E-01
      A(3)=-6.7437328E-02
      A(4)=-8.7753624E-01
      A(5)= 1.4936511E-01
      A(6)=-1.9749393E-01
      A(7)= 4.1687626E-01
      A(8)= 3.8195432E-01
      A(9)=-1.4141054E-01
      A(10)=-2.9439950E-02
      A(11)=1.1908339E-01

      Teff0=(Teff-10000.0)/1000.00
      logg0=(logg-8.00000)/1.00000
      Shift=A(1)+A(2)*exp(-(A(3)+(A(5)+A(7)*exp(-A(8)*
     *((Teff0-A(9))**2)))*Teff0+(A(6)+A(10)*Teff0+A(11)*logg0)*
     *logg0)**2*((Teff0-A(4))**2))
      
      ML17_to_3D_dlogg = Shift
      return
      end 
\end{lstlisting}
~\\

\end{appendix}

\begin{appendix} 

\section{Correction functions for IDL}

\subsection{1D ML2/$\alpha$ = 0.8 to 3D $T_{\rm eff}$ corrections}

\lstset{
    breaklines     = true,
    numbers        = left,
    stepnumber     = 1,
}
\lstset{language=IDL}
\begin{lstlisting}
function ML18_to_3D_dTeff, Teff, logg
;
; IN: Teff, effective temperature (K)
; logg, surface gravity (g in cm/s2)
;
; OUT: Teff correction (K)
; Fri Jun 14 20:16:39 2013 B.F.

A=dindgen(8)
A(00)=1.0947335E-03
A(01)=-1.8716231E-01
A(02)=1.9350009E-02
A(03)=6.4821613E-01
A(04)=-2.2863187E-01
A(05)=5.8699232E-01
A(06)=-1.0729871E-01
A(07)=1.1009070E-01

Teff0=(Teff-10000.0)/1000.00
logg0=(logg-8.00000)/1.00000
Shift=A(00)+(A(01)+A(06)*Teff0+A(07)*logg0)*exp(-(A(02)+A(04)*Teff0+A(05)*logg0)^2*((Teff0-A(03))^2))

return, Shift*1000.00+0.00000

end
\end{lstlisting}
~\\

\subsection{1D ML2/$\alpha$ = 0.8 to 3D $\log g$ corrections}

\begin{lstlisting}
function ML18_to_3D_dlogg, Teff, logg
;
; IN: Teff, effective temperature (K)
; logg, surface gravity (g in cm/s2)
;
; OUT: log g correction (g in cm/s2)
; Fri Jun 14 21:09:56 2013 B.F.

A=dindgen(12)
A(00)=7.5209868E-04
A(01)=-9.2086619E-01
A(02)=3.1253746E-01
A(03)=-1.0348176E+01
A(04)=6.5854716E-01
A(05)=4.2849862E−01
A(06)=-8.8982873E-02
A(07)=1.0199718E+01
A(08)=4.9277883E-02
A(09)=-8.6543477E-01
A(10)=3.6232756E-03
A(11)=-5.8729354E-02

Teff0=(Teff-10000.0)/1000.00
logg0=(logg-8.00000)/1.00000
Shift=(A(00)+A(04)*exp(-A(05)*((Teff0-A(06))^2)))+A(01)*exp(-A(02)*((Teff0-(A(03)+A(07)*exp(-(A(08)+A(10)*Teff0+A(11)*logg0)^2*((Teff0-A(09))^2))))^2))

return, Shift

end
\end{lstlisting}
~\\

\subsection{1D ML2/$\alpha$ = 0.7 to 3D $T_{\rm eff}$ corrections}

\begin{lstlisting}
function ML17_to_3D_dTeff, Teff, logg
;
; IN: Teff, effective temperature (K)
; logg, surface gravity (g in cm/s2)
;
; OUT: Teff correction (K)
; Tue Jun 18 18:58:49 2013 B.F.

A=dindgen(8)
A(00)=-1.0461690E-03
A(01)=-2.6846737E-01
A(02)=3.0654611E-01
A(03)=1.8025848E+00
A(04)=1.5006909E-01
A(05)=1.0125295E-01
A(06)=-5.2933335E-02
A(07)=-1.3414353E-01

Teff0=(Teff-10000.0)/1000.00
logg0=(logg-8.00000)/1.00000
Shift=A(00)+(A(01)+A(04)*Teff0+(A(05)+A(06)*Teff0+A(07)*logg0)*logg0)*exp(-A(02)*((Teff0-A(03))^2))

return, Shift*1000.00+0.00000

end
\end{lstlisting}
~\\

\subsection{1D ML2/$\alpha$ = 0.7 to 3D $\log g$ corrections}

\begin{lstlisting}
function ML17_to_3D_dlogg, Teff, logg
;
; IN: Teff, effective temperature (K)
; logg, surface gravity (g in cm/s2)
;
; OUT: log g correction (g in cm/s2)
; Tue Jun 18 18:56:55 2013 B.F.

A=dindgen(11)
A(00)=1.1922481E-03
A(01)=-2.7230889E-01
A(02)=-6.7437328E-02
A(03)=-8.7753624E-01
A(04)=1.4936511E-01
A(05)=-1.9749393E-01
A(06)=4.1687626E-01
A(07)=3.8195432E-01
A(08)=-1.4141054E-01
A(09)=-2.9439950E-02
A(10)=1.1908339E-01

Teff0=(Teff-10000.0)/1000.00
logg0=(logg-8.00000)/1.00000
Shift=A(00)+A(01)*exp(-(A(02)+(A(04)+A(06)*exp(-A(07)*((Teff0-A(08))^2)))*Teff0+(A(05)+A(09)*Teff0+A(10)*logg0)*logg0)^2*((Teff0-A(03))^2))

return, Shift

end
\end{lstlisting}

\end{appendix}

\end{document}